\documentclass[12pt]{iopart}

\usepackage{amsfonts}
\usepackage{amssymb}
\usepackage{iopams}
\usepackage{setstack}
\usepackage{graphicx}

\newcommand{\sech}{\mathrm{sech}}

\begin{document}

\paper{Traveling waves of the regularized short pulse equation}

\author{Y. Shen$^1$, T.P.\ Horikis$^2$, P.G.\ Kevrekidis$^3$, D.J.\ Frantzeskakis$^4$}
\address{$^1$Department of Mathematical Sciences, University of Texas at Dallas,
Richardson, TX 75080, USA}
\address{$^2$Department of Mathematics, University of
Ioannina, 45110 Ioannina, Greece}
\address{$^3$Department of Mathematics and
Statistics, University of Massachusetts, Amherst MA 01003-4515, USA}
\address{$^4$Department of Physics, University of Athens, Panepistimiopolis,
Zografos, Athens 15784, Greece }

\ead{\mailto{yxs135630@utdallas.edu}, \mailto{horikis@uoi.gr},
\mailto{kevrekid@math.umass.edu}, \mailto{dfrantz@phys.uoa.gr}}

\begin{abstract}
In the present work, we revisit the so-called regularized short pulse
equation (RSPE) and, in particular, explore the traveling wave solutions of
this model. We theoretically analyze and numerically evolve two sets of
such solutions. First, using a fixed point iteration scheme, we numerically
integrate the equation to find solitary waves. It is found that these
solutions are well approximated by a truncated series of hyperbolic
secants. The dependence of the soliton's parameters (height, width, etc) to
the parameters of the equation is also investigated. Second, by developing
a multiple scale reduction of the RSPE to the nonlinear Schr{\"o}dinger
equation, we are able to construct (both standing and traveling) envelope
wave breather type solutions of the former, based on the solitary wave
structures of the latter. Both the regular and the breathing traveling wave
solutions identified are found to be robust and should thus be amenable to
observations in the form of few optical cycle pulses.
\end{abstract}

\pacs{42.65.Tg, 42.81.Dp, 05.45.Yv, 02.30.Jr}

\submitto{\JPA}

\section{Introduction}

Over the last decade, there has been an intense interest in the study of
ultrashort pulses with a duration of a few optical cycles, due to their
numerous applications in contexts including, but not limited to, nonlinear
optics, attosecond physics, light-matter interactions and harmonic
generation~\cite{bk}. Nonlinear media present an especially interesting
setting for the propagation of such pulses due to their intensity-dependent
refractive index. As a result, there is a continuously expanding volume of
literature attempting to model and systematically explore such pulses using a
diverse array of nonlinear wave equations, including in one-dimension (1D)
the modified Korteweg-de Vries (mKdV) equation \cite{mih1}, the sine-Gordon
(sG) equation \cite{ls,mih2}, and combined mKdV-sG equations
\cite{mih3,mih4,LM} (see also the recent review \cite{mihrev}). Relevant
models have also been developed by adapting to special settings such as,
e.g., near the zero dispersion frequency~\cite{vladimir}. In the
two-dimensional case, relevant generalizations utilize  the generalized
Kadomtsev-Petviashvilli equation \cite{mih5} and explore even features such
as the collapse of ultrashort spatiotemporal pulses \cite{LKM}.

A considerable parallel physical as well as mathematical activity has also
been focused on a different class of models that was initiated by the work of
Ref.~\cite{sw}. There, and in the context of nonlinear fiber optics, the
so-called short-pulse equation (SPE) was derived from Maxwell's equations.
Importantly, it was shown that results obtained in the framework of the SPE
model compare favorably to ones pertaining to the original Maxwell's
equations~\cite{sw,sw2}, and are more accurate than results corresponding to
the nonlinear Schr{\"o}dinger equation (NLS), that is traditionally used in
this (nonlinear fiber optics) context. This effort spurred numerous further
studies. These were in part due to the illustrated relevance of the SPE as a
suitable model, e.g., in nonlinear left-handed metamaterials~\cite{tsitsas2}
from the physical perspective. However, numerous interesting conclusions also
arose from the mathematical setting due to the complete integrability of the
model, and its intimate connection to the sG equation \cite{saksak1,yao}, the
derivation of its infinite hierarchy of conservation laws~\cite{brun}, as
well as the identification of loop solitons, breathers and numerous other
(including periodic) solutions that were identified
therein~\cite{tsitsas2,horikis,saksak2,spesol,mats}. Such SPE models were
also developed in higher-dimensional settings, cf.~\cite{shen2d} and
integrable discretizations thereof were obtained as well~\cite{maruno}.

Recently, a number of variant SPE models have also appeared. Arguably, the
most well known example of this kind is the so-called regularized short pulse
equation (RSPE) originally introduced in~\cite{cmj}, again in the nonlinear
fiber optics context. In that work, the regularization was argued as arising
from the next term in the expansion of the susceptibility (within Maxwell's
equations). Importantly, it was shown that while the SPE does not support
traveling wave (TW) solutions (at least in the class of piecewise smooth
functions with one discontinuity), the RSPE model does exhibit such solutions
(under certain conditions for the coefficients of the equation), whose
existence was established via Fenichel theory and a Melnikov type argument.
Extending this work, in~\cite{mcjs}, the existence of certain types of
multi-pulse traveling waves was also shown by means of a geometric integral
condition. We should note in passing here that other more complex
higher-order correction models to the SPE have been proposed recently as,
e.g., in the work of~\cite{chung}.

In the present work, the focus is on the RSPE model, considered in the
following form:
\begin{equation}
u_{xt} + \alpha u+ (u^3)_{xx} + \beta u_{xxxx} = 0,
\label{rspe}
\end{equation}
where $u(x,t)$ is the unknown field, subscripts denote partial derivatives,
while $\alpha$ and $\beta$ are real parameters (here, we use a notation
similar to that of Ref.~\cite{cmj} and keep both parameters --although one of
them can be scaled out up to its sign). Our aim is to explore traveling wave
solutions of Eq.~(\ref{rspe}), which we investigate through a combination of
analytical and numerical techniques. In particular, we actually explore two
classes of solutions. On one hand, revisiting the proposal (and proof)
of~\cite{cmj} about the existence of traveling waves of the RSPE that cannot
be present in the SPE, we identify such solutions numerically using a fixed
point iteration scheme (briefly discussed in Appendix A). We also show that
such solutions can be well approximated by a truncated series of hyperbolic
secant functions. On the other hand, we revisit the --arguably-- most robust
structures featured by the SPE (without the regularization), namely the
standing and traveling wave breathers. By utilizing a reduction of the RSPE
in the appropriate space and time scales (i.e., through a multi-scale
expansion), we are able to revert the model to the standard NLS form. By then
employing the NLS standing wave solitons, we are able to reconstruct
breathers of the RSPE both of the bright and dark type (the latter only exist
in the case of periodic boundary conditions). Our numerical simulations
suggest the structural robustness of {\it both} classes of solutions, with
the traveling waves propagating undistorted (despite the addition of small
random perturbations), and with standing or traveling envelope breathers
staying proximal to their NLS analogs. Hence, both classes of solutions
should be possible to observe in the few optical cycle setting.

Our presentation is structured as follows. In section 2, we identify and
numerically characterize (as a function of the corresponding parameters) the
traveling wave (without breathing) of the RSPE. In section 3, we present the
reduction of the RSPE to the NLS and reconstruct the breather solution of the
former from the bright solitons of the latter. Finally, in section 4, we
summarize our findings and discuss a number of directions for future studies.
Some technical aspects of the analysis and computations are discussed in two
appendices.

\section{Solitary waves of the RSPE}

We consider, at first, the solutions identified rigorously through the work
of~\cite{cmj}, namely the traveling solitary waves of Eq.~(\ref{rspe}). These
are identified by going to the co-traveling frame using $\xi=x-ct$ (where $c$
is the velocity), in which the RSPE becomes:
\begin{equation}
-c u''+\alpha u+(u^3)''+\beta u''''=0,
\label{rspe2}
\end{equation}
where primes denote derivatives with respect to $\xi$. Recall \cite{cmj} that
these solutions exist when:
\begin{equation}
\mathrm{sgn}(c)=\mathrm{sgn}(\alpha)=\mathrm{sgn}(\beta).
\label{cond}
\end{equation}
Following the procedure of Refs.~\cite{sprz1,sprz2}, we obtain numerically
the pertinent solitary wave for $c=\alpha=\beta=1$, by means of the spectral
renormalization method. The details of this computation for our case are
summarized in Appendix A. In Fig.~\ref{soliton_rspe} we depict this solution
and its evolution according to Eq. (\ref{rspe}). The evolution was performed
using a fourth order Runge-Kutta algorithm.

\begin{figure}[!htbp]
    \centering
    \includegraphics[width=7cm]{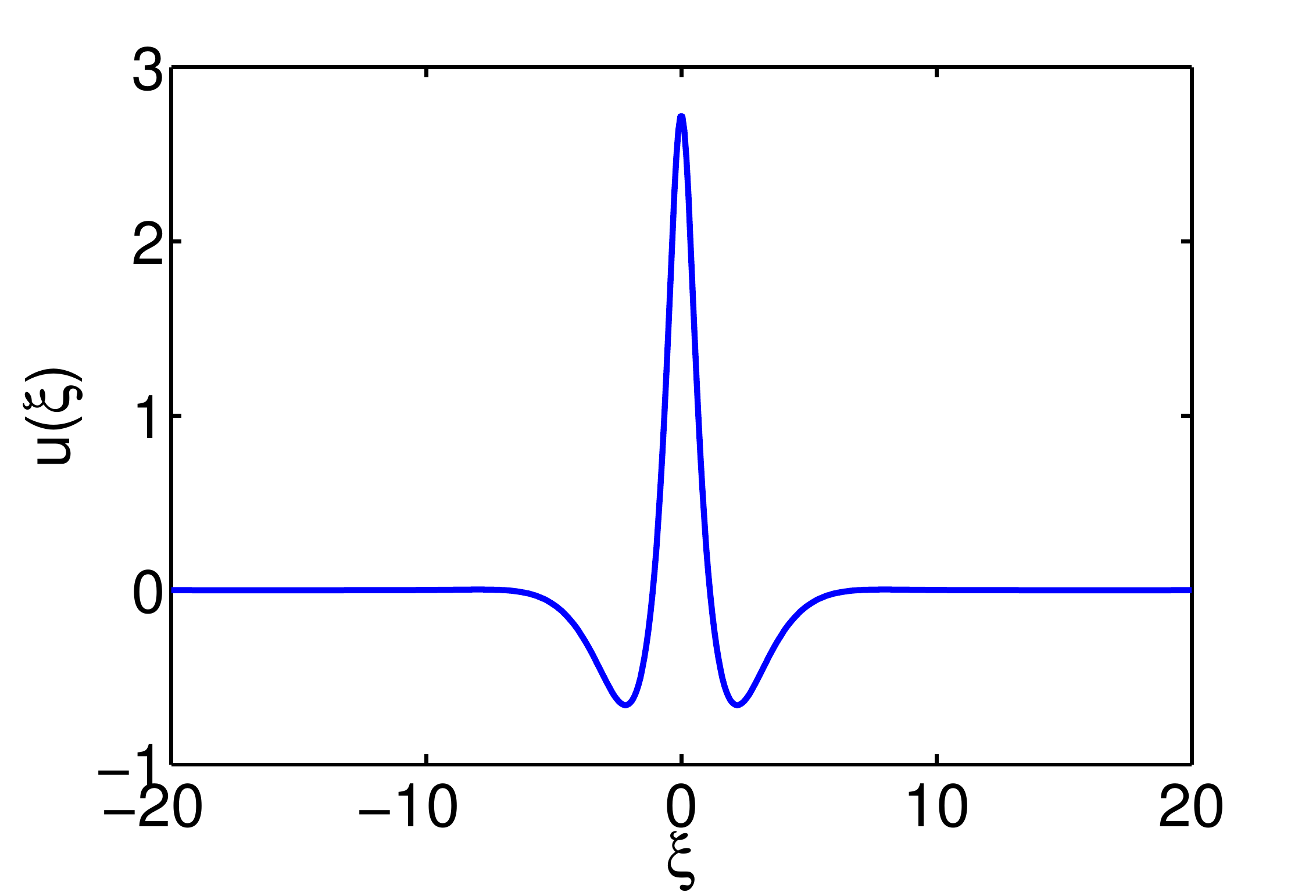}
    \includegraphics[width=8cm]{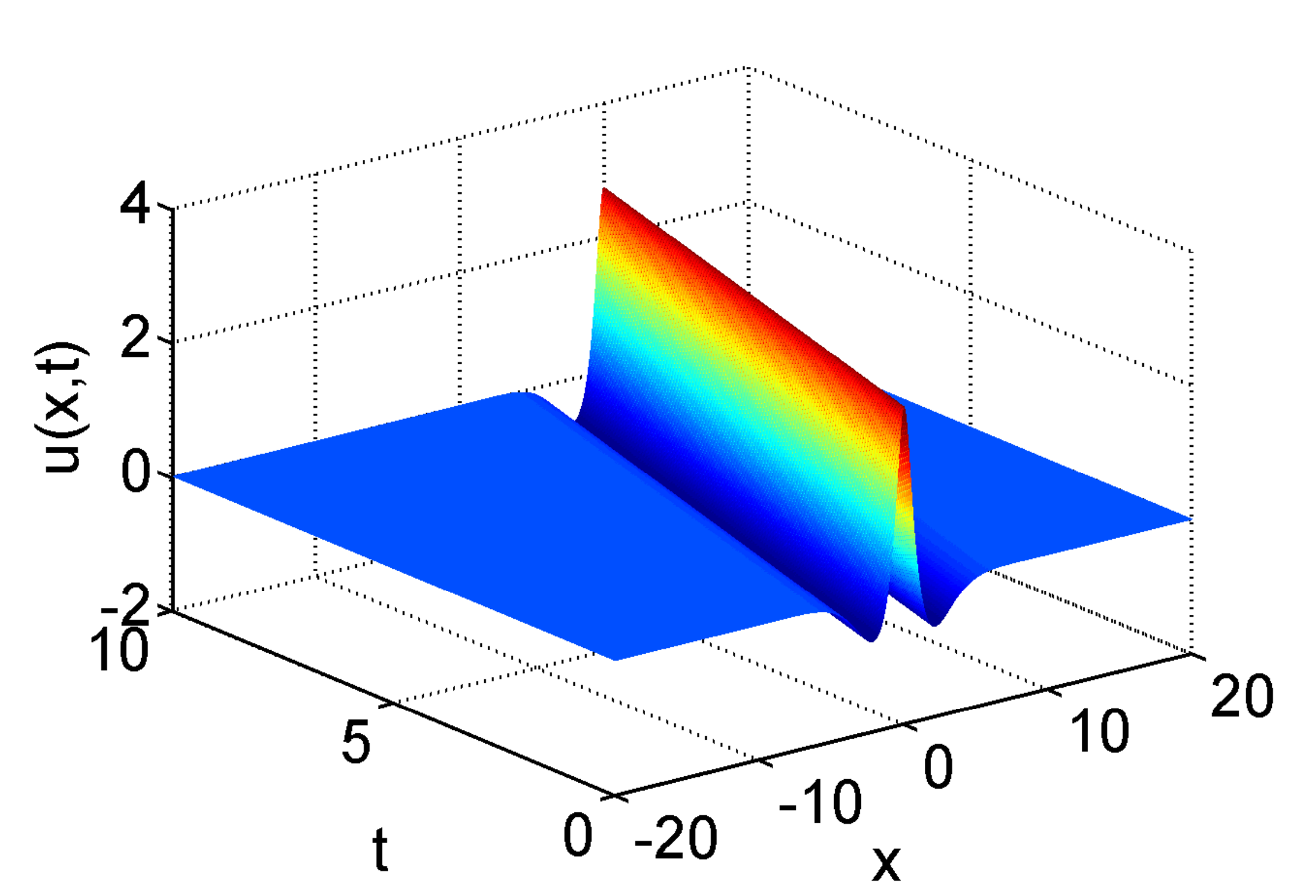}
    \caption{(Color online) The solitary wave (left) and its evolution (right)
    under the RSPE for $c=\alpha=\beta=1$.}
    \label{soliton_rspe}
\end{figure}

\begin{figure}[!htbp]
    \centering
    \includegraphics[width=8cm]{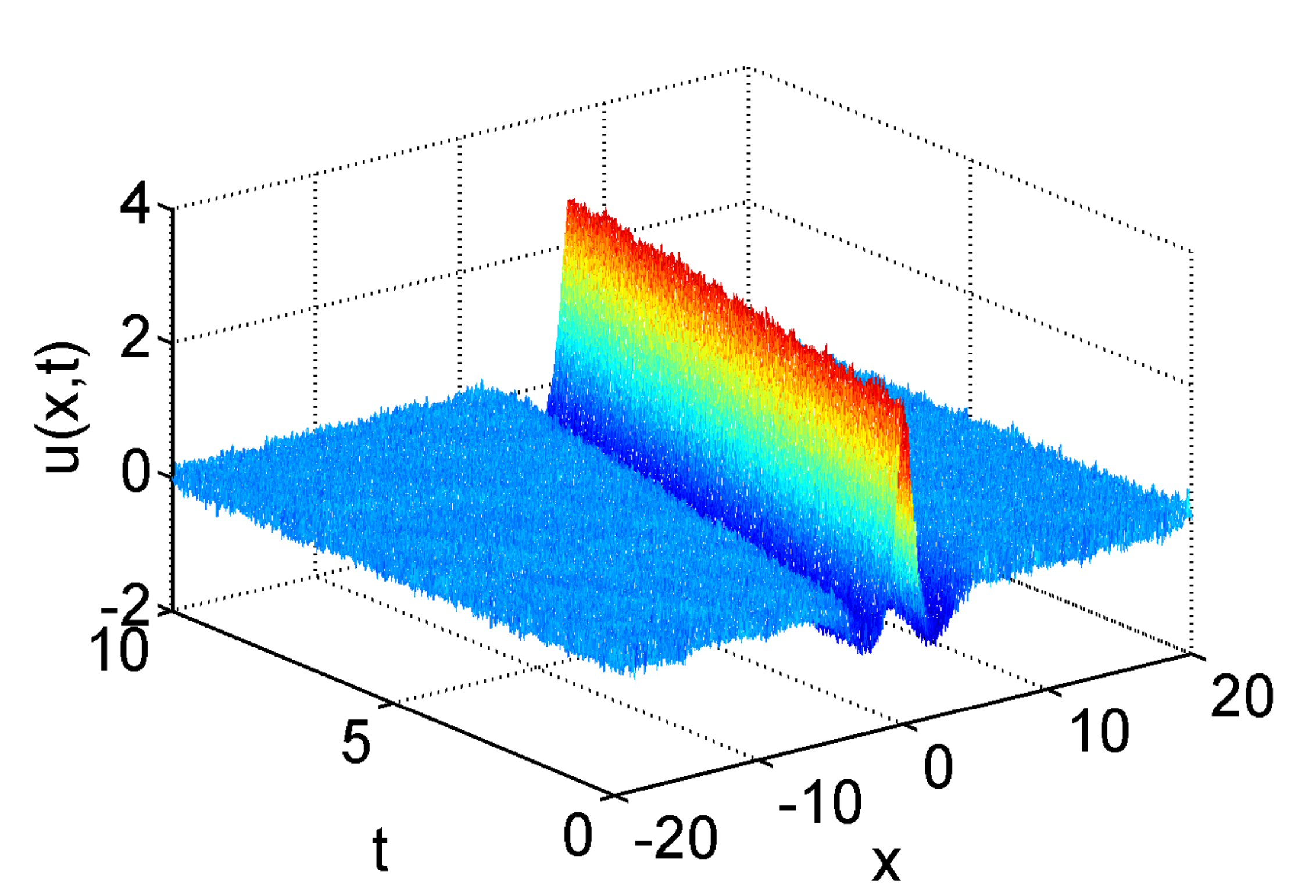}
    \caption{(Color online) The solitary wave of Fig. \ref{soliton_rspe} evolved with added noise.}
    \label{noise.fig}
\end{figure}

To test the stability of this result we evolve again the same solitary wave
but with 10\% initial noise added. The resulting evolution is shown in
Fig.~\ref{noise.fig}. The figure is suggestive of the robustness of the
solitary wave given the preservation of its traveling characteristics under
the addition of noise. We note in passing that other simulations
corresponding to different parameter values have led to similar results,
namely indicating robustness of the traveling wave.

\begin{figure}[!htbp]
    \centering
    \includegraphics[width=5cm]{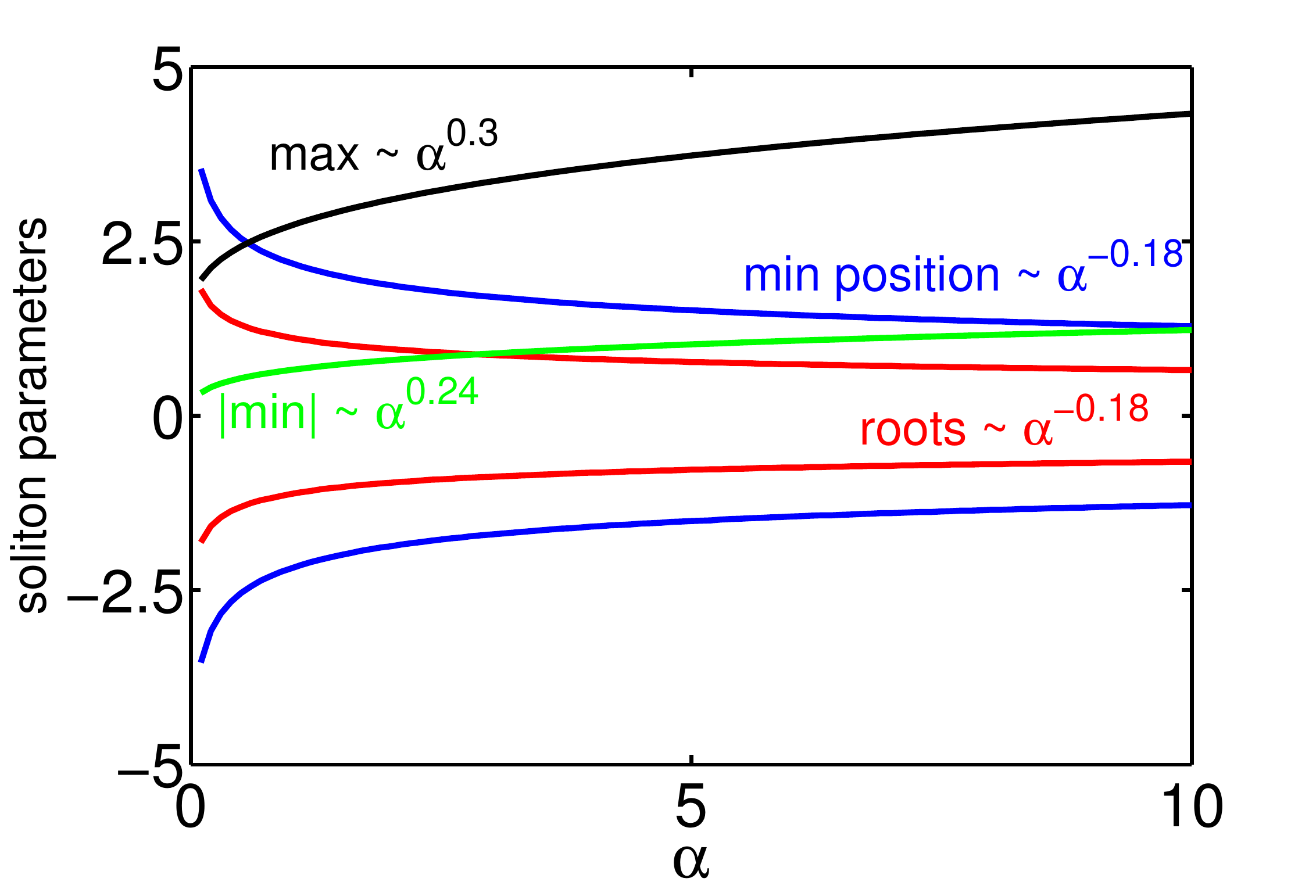}
    \includegraphics[width=5cm]{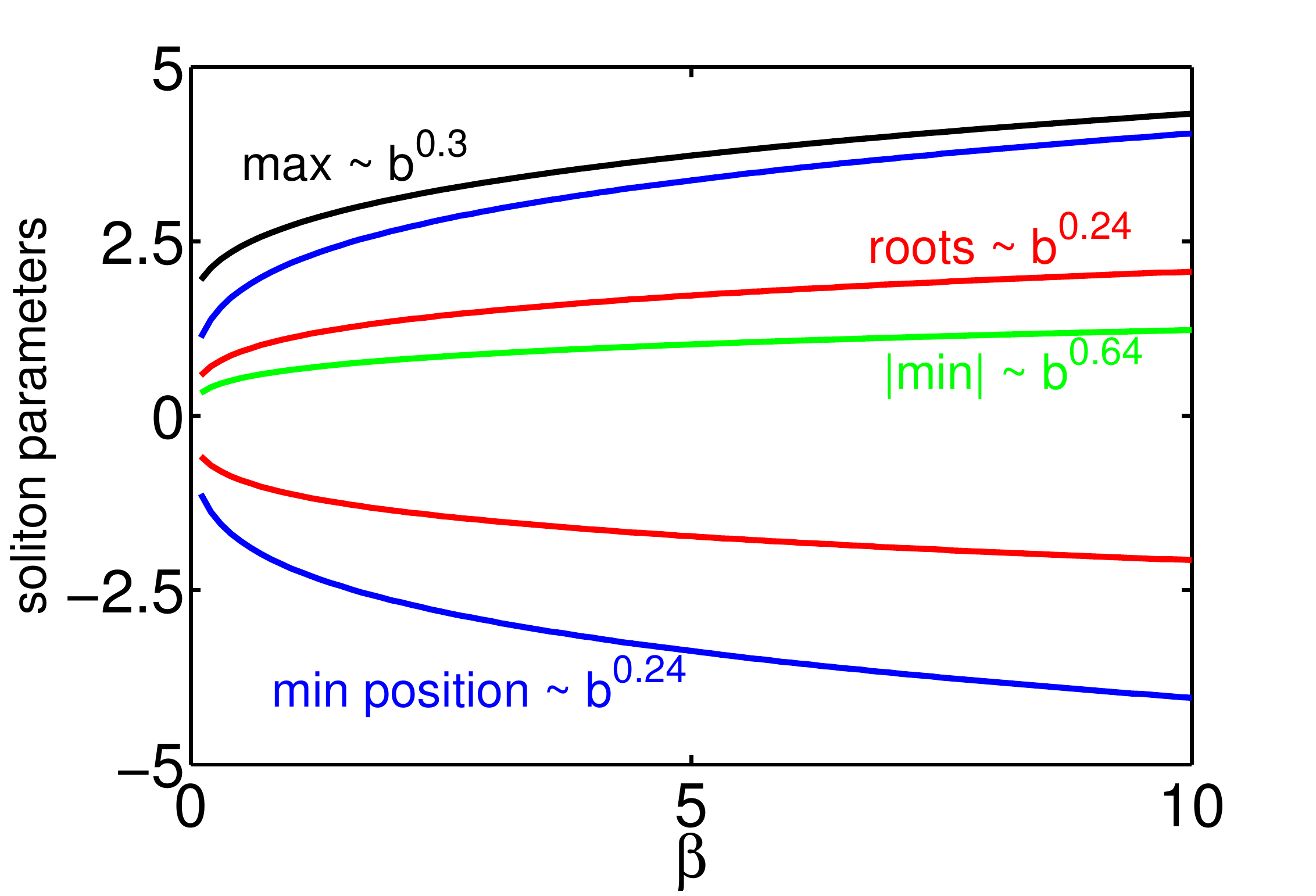}
    \includegraphics[width=5cm]{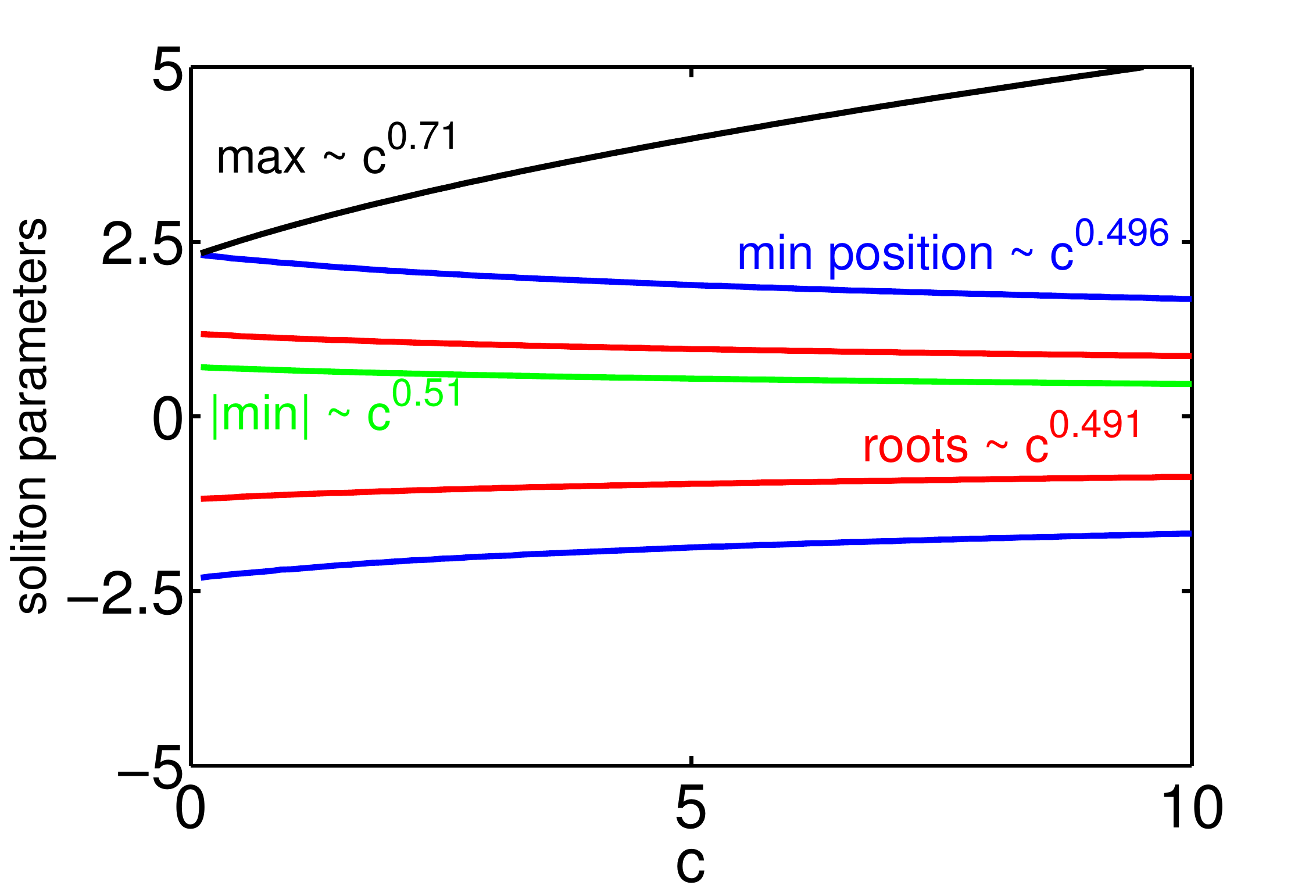}
    \caption{(Color online) The change of the soliton parameters with $\alpha$, $\beta$ and $c$.}
    \label{fig.param}
\end{figure}

Next, we examine the behaviour of the solution when parameters $\beta$,
$\alpha$ and $c$ change. To do this, every time a parameter changes while the
other two are kept constant (equal to 1), and the solution's characteristics
values are measured i.e., we perform mono-parametric continuations. The
related curves are fitted to power laws; the results are depicted in
Fig.~\ref{fig.param}. The figures depict the change of the soliton's height,
width (identified as the position where the soliton crosses the $x$-axis,
i.e., the location of the corresponding root), minimum and the position of
the minimum. The complete fitted curves are summarized for all parameters in
Table~\ref{table.param}.

\begin{table}[!htbp]
\centering
\caption{\label{table.param}The change of the soliton parameters according to the RSPE parameters.}
\begin{indented}
\item[
\hspace{-2cm}
\begin{tabular}{@{}llll}\br
Soliton parameter & $\alpha$ & $\beta$ & $c$ \\ \mr
Max &  $1.578 \alpha^{0.3045}+1.154$ & $1.578 \beta^{0.3045}+1.154$ & $0.5724 c^{0.711}+2.17$ \\
Min & $0.7848 \alpha^{0.2376}-0.1254$ & $0.8687 \beta^{0.6419}-0.9971$ & $-0.0877 c^{0.5084}+0.7433$ \\
Root & $1.355 \alpha^{-0.1817}-0.2368$ & $1.267 \beta^{0.2428}-0.1475$ & $-0.1193 c^{0.4963}+1.234$ \\
Position of min & $2.681 \alpha^{-0.1792}-0.4937$ & $2.515 \beta^{-0.2402}-0.3246$ & $0.2418
c^{0.4915}-2.416$ \\
\br
\end{tabular}]
\end{indented}
\end{table}

Two particular limits are of interest here: $\alpha=0$ and $\beta=0$. For the
first, one can observe that the amplitude of the solution slowly grows from
the mKdV limit of $\alpha=0$ (which is the starting point in our
continuations), while concurrently a root emerges. In particular, the
monotonic soliton of the mKdV (i.e., a ${\rm sech}$ traveling pulse) develops
a shelf, which is entirely absent in the mKdV limit, and features a root
which scales as a negative power of $\alpha$. This means that the region of
uniform sign continuously shrinks at the center as $\alpha$ is increased.
Moreover, the height of this shelf also continuously increases as a function
of $\alpha$. This is consistent with the observations of Ref.~\cite{kodama}
for the case of small values of $\alpha$. The second limit, $\beta=0$, is
more complicated as there are no traveling waves for the SPE. An attempt to
perturbatively explain this limit (although not for the RPSE case) was made
in Ref.~\cite{spe.pert} using a variational approach since even linearization
of the exact solution (of the SPE) is not tractable.

\begin{figure}[!htbp]
    \centering
    \includegraphics[width=8cm]{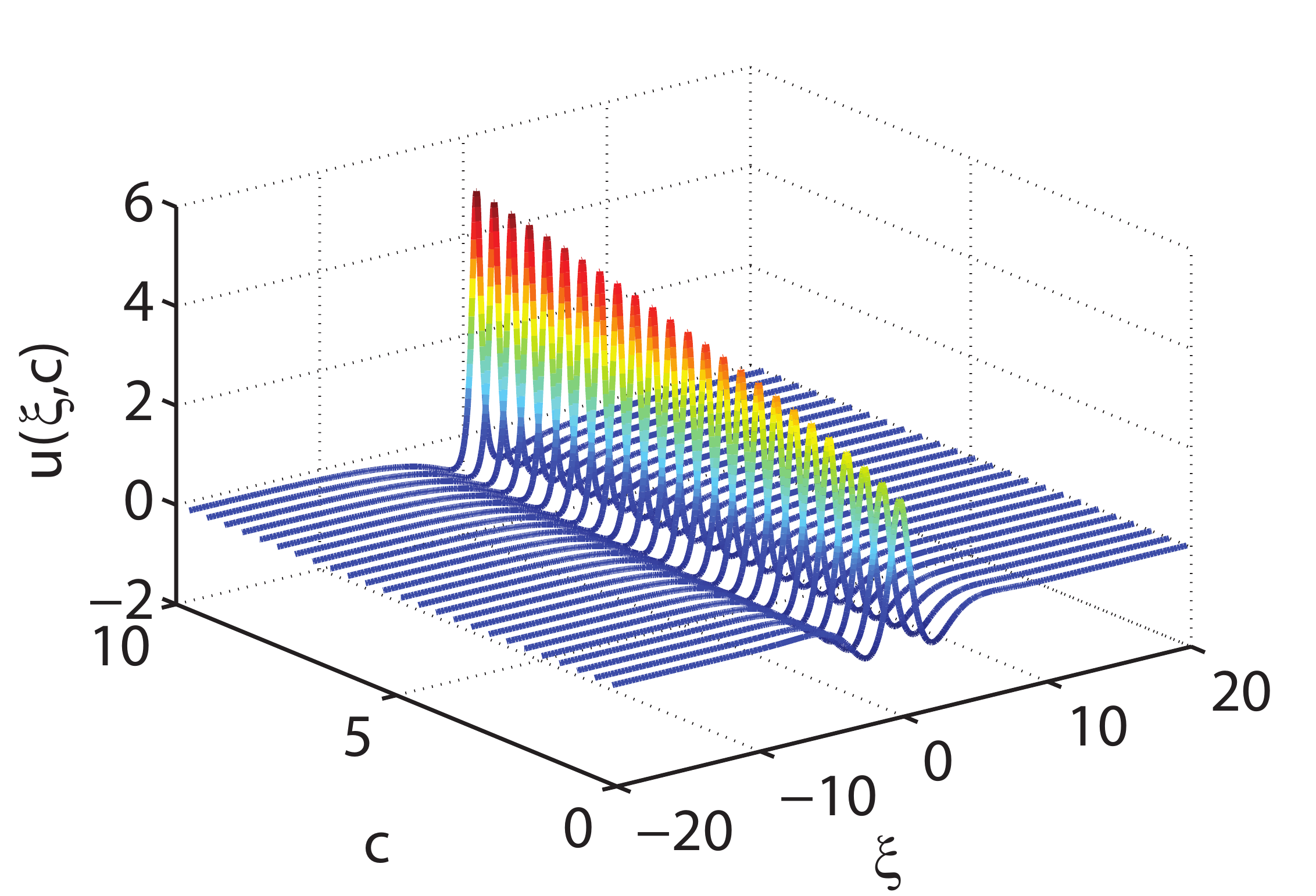}
    \caption{(Color online) The solution of the RSPE for different values of $c$.}
    \label{fig.c}
\end{figure}

Finally, we observe a monotonic growth of the solution amplitude as $c$ is
increased. On the other hand, the region associated with the central part of
the solution (i.e., up to the location of the root) is very weakly affected
by the variation of the speed $c$. For illustrative purposes we show in
Fig.~\ref{fig.c} the change of the solution with the speed $c$.

\begin{figure}[!htbp]
    \centering
    \includegraphics[width=7cm]{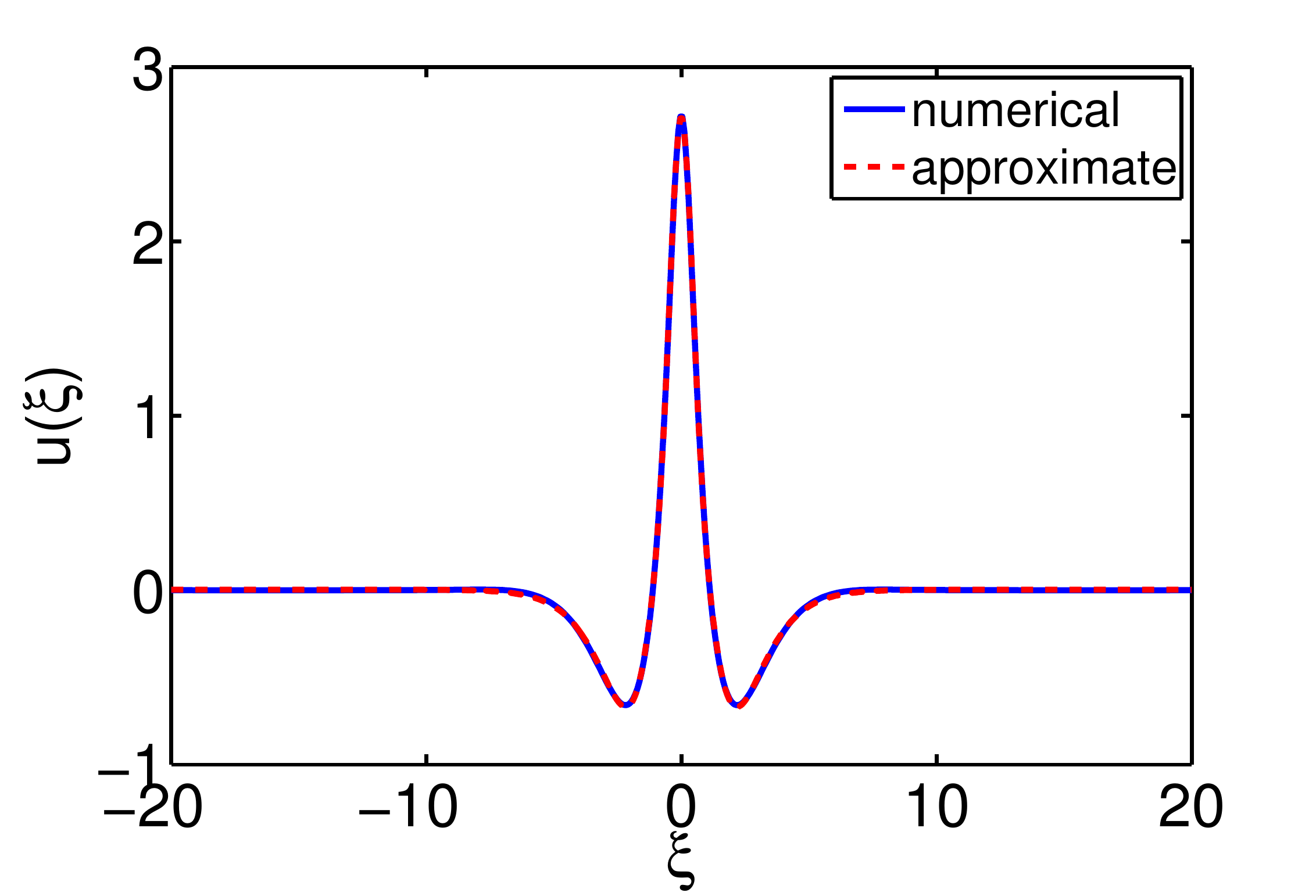}
    \includegraphics[width=8cm]{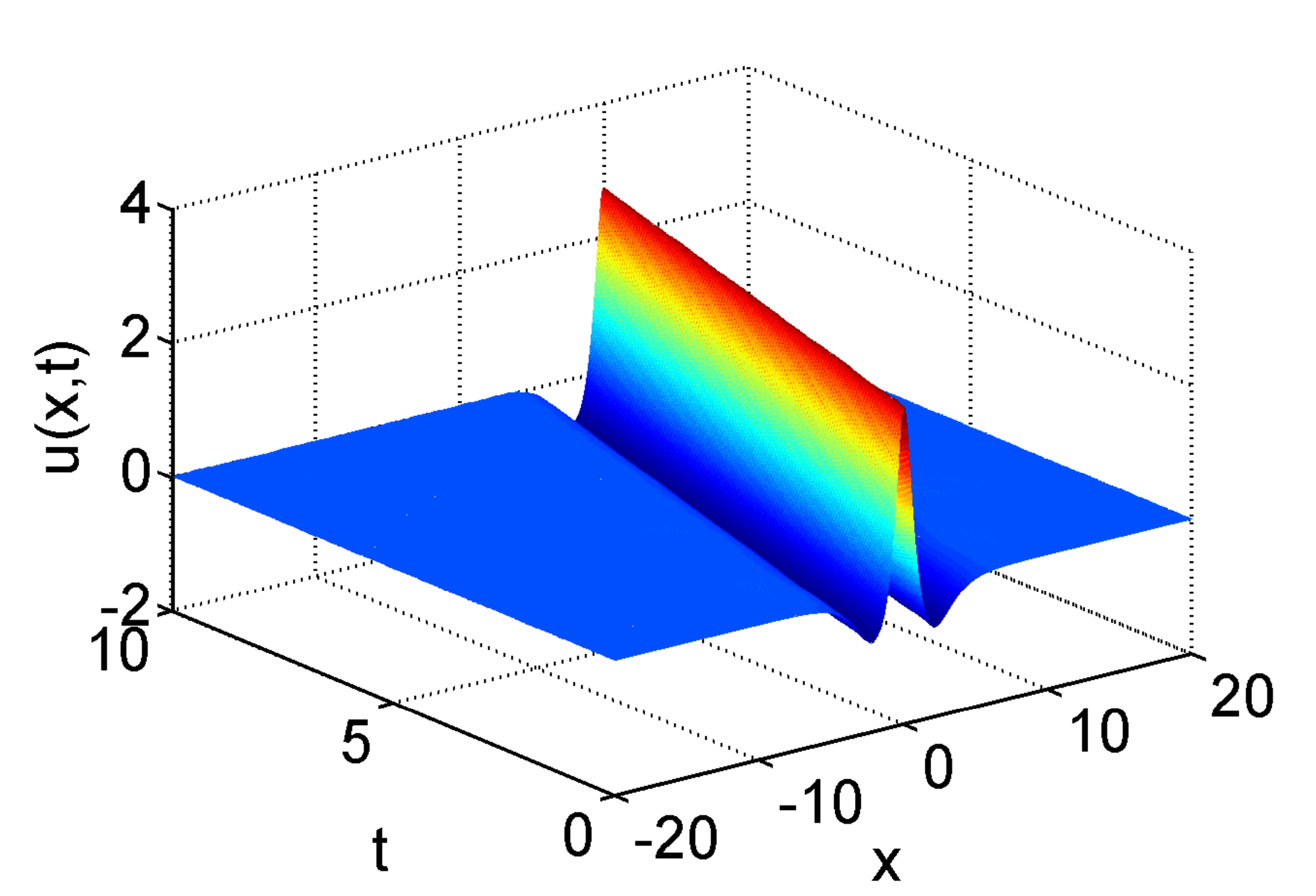}
    \caption{(Color online) Left: The comparison between the numerically obtained solution
    and the analytical approximation of Eq.~(\ref{apr}).
    Right: The evolution of the approximate solution.}
    \label{compare}
\end{figure}

It is also useful to develop an analytical expression for the form of the
solution. In order to do so, we can approximate the traveling wave, to
excellent accuracy, with a truncated series of $\mathrm{sech}$'s according
to:
\begin{equation}
u(\xi)= a_0 +a_1\mathrm{sech}(b\xi)+a_2\mathrm{sech}^2(b\xi)+a_3\mathrm{sech}^3(b\xi)
+a_4\mathrm{sech}^4(b\xi)
\label{apr}
\end{equation}
The comparison between the numerically found solution and the approximation
of Eq.~(\ref{apr}) is shown in Fig.~\ref{compare}. The figure also presents
the propagation of an initial condition in the form of Eq.~(\ref{apr}). It is
evident that the approximate solution provides a very accurate description of
the dynamics, maintaining the robust traveling characteristics of the
corresponding exact solution.

The exact parameters for this solution read
\[
a_0=0.002,\; a_1=-6.952,\; a_2=23.17,\; a_3=-27.88,\; a_4=14.37,\; b=0.986,
\]
while the coefficient of determination is found to be $r^2=0.999$.

\section{Approximate envelope solitary waves of the RSPE}

\subsection{Multi-scale analysis and the connection to NLS}

In this section we identify standing and traveling breathers of the RSPE
model. Such solutions, which are reminiscent of the breather solutions of the
regular SPE model (see, e.g., Refs.~\cite{tsitsas2,spesol,mats}), can be
found by reducing the RSPE to the NLS equation by applying a formal
multiscale expansion method. In particular we assume that the unknown field
$u$ is a function of multiple spatial and temporal scales
$(x_0,x_1,t_0,t_1,t_2)$, defined as
\begin{eqnarray*}
&&x_0 = x,\ \  x_1 = \epsilon x \\
&&t_0 = t,\ \  t_1 = \epsilon t,\ \  t_2 = \epsilon^2 t,
\end{eqnarray*}
where $\epsilon$ is a formal small parameter. Furthermore, we expand $u$ as:
\begin{eqnarray}
\label{expansion}
u = \epsilon u_1 + \epsilon^2 u_2 + \epsilon^3 u_3 +...
\end{eqnarray}
Then, to leading order, $O(\epsilon)$, we obtain the linear equation
\begin{eqnarray}
\partial_{x_0}\partial_{t_0} u_1+\alpha u_1 +\beta \partial_{x_0}^4u_1 = 0
\label{e0}
\end{eqnarray}
whose solution is sought in the form
\[
u_1 = A(x_1,t_1,t_2)e^{i\theta} + A^*(x_1,t_1,t_2)e^{-i\theta}
\]
where $\theta = kx_0-\omega t_0$ and $k\omega + \beta k^4 + \alpha = 0$ and
$*$ denotes complex conjugate. The phase velocity is as usual defined as
$c_p=\omega/k$. For $u_2$ to $O(\epsilon^2)$, we have:
\begin{eqnarray}
\partial_{x_0}\partial_{t_0} u_2+\alpha u_2 +\beta \partial_{x_0}^4u_2 = -[\partial_{t_0}\partial_{x_1}
u_1+\partial_{x_0}\partial_{t_1} u_1 +4\beta \partial_{x_0}^3\partial_{x_1}u_1]
\label{e2}
\end{eqnarray}
Cancelling secular terms from the right hand side of Eq.~(\ref{e2}), we
obtain
\begin{eqnarray}
k\partial_{t_1}A - (4\beta k^3+\omega)\partial_{x_1}A=0.
\label{zero2}
\end{eqnarray}
We now let $\xi = x_1-c_g t_1$, $t_2 = \tau$, and
$A(x_1,t_1,t_2) = A(\xi, \tau)$. Then, from
Eq.~(\ref{zero2}) we get the group velocity
\begin{eqnarray}
c_g = -3k^2\beta+\frac{\alpha}{k^2}= \frac{d\omega}{dk} .
\label{cg}
\end{eqnarray}
Additionally, Eq.~(\ref{e2}) is now homogeneous and thus has the solution
$u_2= 0$. Finally, at $O(\epsilon^3)$, we obtain
\begin{eqnarray*}
\partial_{x_0}\partial_{t_0} u_3+\alpha u_3 +\beta \partial_{x_0}^4u_3 =\\
-[\partial_{x_1}\partial_{t_1}
u_1+\partial_{x_0}\partial_{t_2} u_1 +6u_1(\partial_{x_0}u_1)^2+3u_1^2\partial_{x_0}^2 u_1+
6\beta\partial_{x_0}^2\partial_{x_1}^2u_1].
\end{eqnarray*}
Cancelation of the secular terms yields the NLS model
\begin{eqnarray}
ik\partial_{t_2}A +\partial_{x_1}\partial_{t_1}A- 6\beta k^2\partial_{x_1}^2A-3k^2|A|^2A=0;
\label{zero3}
\end{eqnarray}
while the third order equation reduces to
\begin{eqnarray*}
\quad \partial_{x_0}\partial_{t_0} u_3+\alpha u_3 +\beta \partial_{x_0}^4u_3 =
-9k^2A^3e^{3i\theta}-9k^2{A^*}^3e^{-3i\theta}
\end{eqnarray*}
with $u_3 = Be^{3i\theta}+B^*e^{-3i\theta}$,
where $B = \frac{-9k^2}{81\beta k^4+9\omega k +\alpha}A^3$.

Concluding, Eq.~(\ref{zero3}) leads in a self-consistent
fashion through the above multi-scale analysis to a NLS model for
the dynamics in the form of:
\begin{eqnarray}
i\partial_{\tau}A +P\partial_{\xi}^2A + Q|A|^2A=0,
\label{NLS}
\end{eqnarray}
where the dispersion and nonlinearity coefficients, $P$ and $Q$, respectively, are given by:
\[
P= \frac{1}{2} \frac{\partial^2 \omega}{\partial k^2} = -3k\beta - \frac{\alpha}{k^3}, \qquad Q= -3k.
\]
Based on the above prescription, for given parameters $\alpha$ and $\beta$,
we can select a particular wavenumber $k$, which, in turn, gives rise to a
frequency $\omega$ [through Eq.~(\ref{e0})] and a group velocity [through
Eq.~(\ref{cg})]. Then the exact analytical soliton solution [in the form of
either a bright (for $PQ>0$) or a dark soliton (for $PQ<0$) --see below] of
the NLS Eq.~(\ref{NLS}) for a given choice of $\epsilon$ can be used to
reconstruct the first few orders of the solution of the RSPE, namely $u_1$
and $u_3$ (recall that $u_2=0$). Recombining these in our series expansion of
Eq.~(\ref{expansion}), we are able to reconstruct an accurate approximation
of a breather of the RSPE. A similar approximation was used in
Ref.~\cite{kutz} in the context of mode-locked lasers.

We note in passing that it is also possible to derive a higher-order NLS
(HNLS) equation, which also admits solitary wave solutions (cf. Appendix B).
Nevertheless, direct numerical simulations that will be presented below show
that solutions of the RSPE can already be described fairly accurately by
respective soliton solutions of the regular NLS, Eq.~(\ref{NLS}), hence we
restrict our numerical considerations to the latter without resorting to the
HNLS below.

\subsection{Bright Breathers}

First, we consider moving breathers of the RSPE model, as described by the focusing
NLS Eq.~(\ref{NLS}) for $PQ>0$. In this case, the NLS admits the bright soliton solution
\[
A(\xi,\tau) = \sqrt{\frac{2}{|Q|}}\sech\left(\frac{1}{\sqrt{|P|}}\xi\right)\exp(i\tau).
\]
\begin{figure}[!htbp]
    \centering
	\includegraphics[width=.4\textwidth]{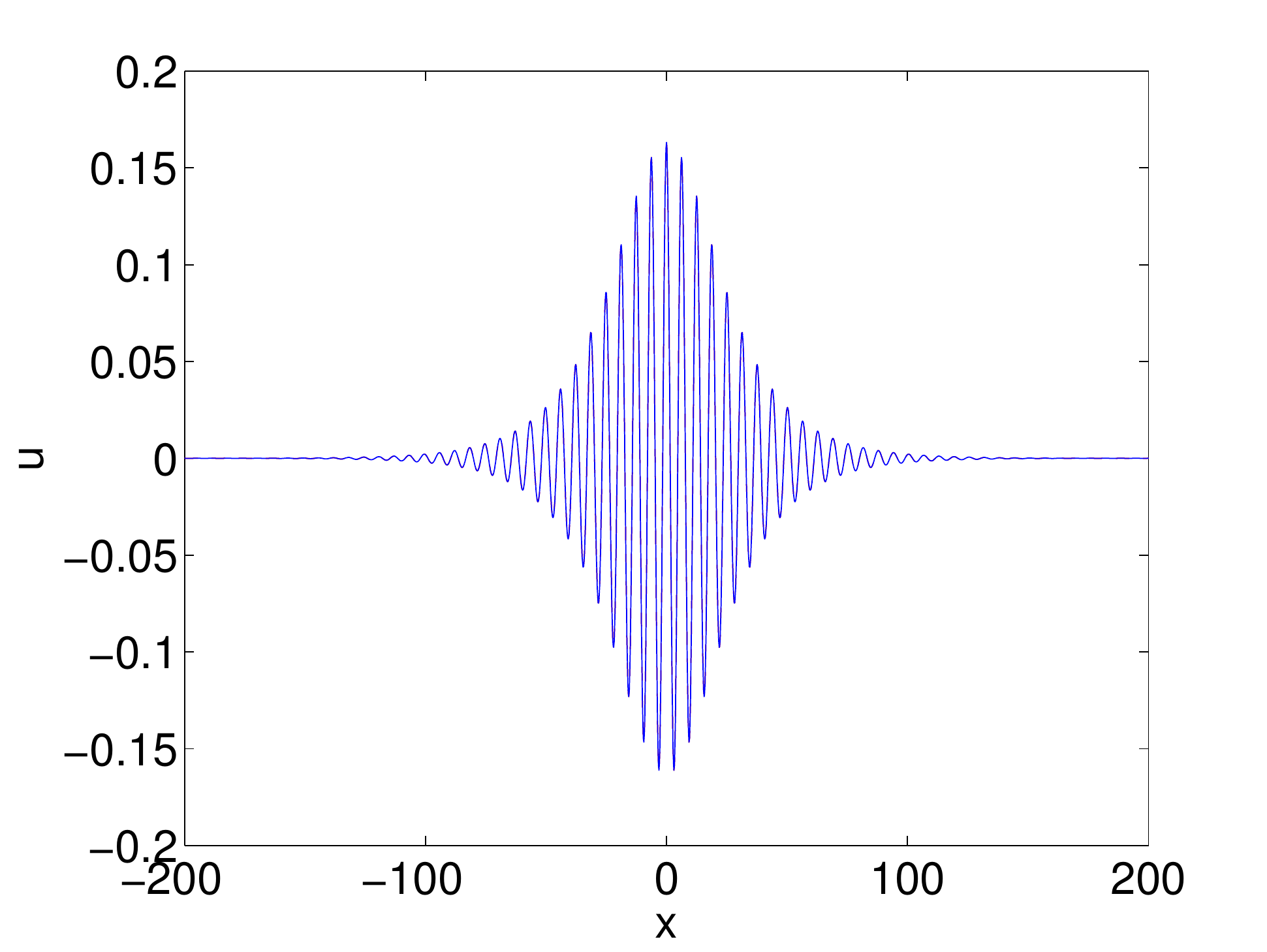}
	\includegraphics[width=.4\textwidth]{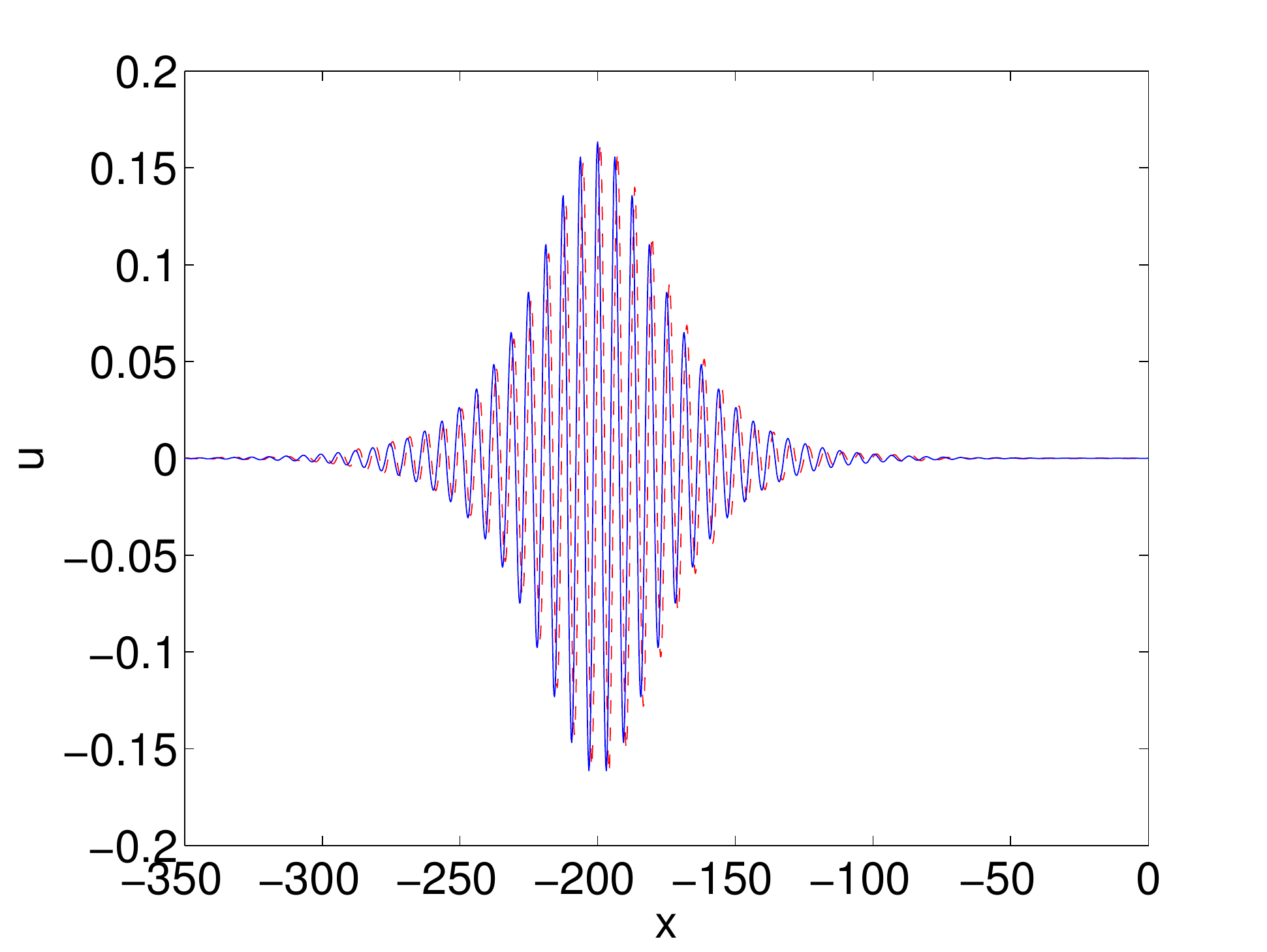}
	\includegraphics[width=.32\textwidth]{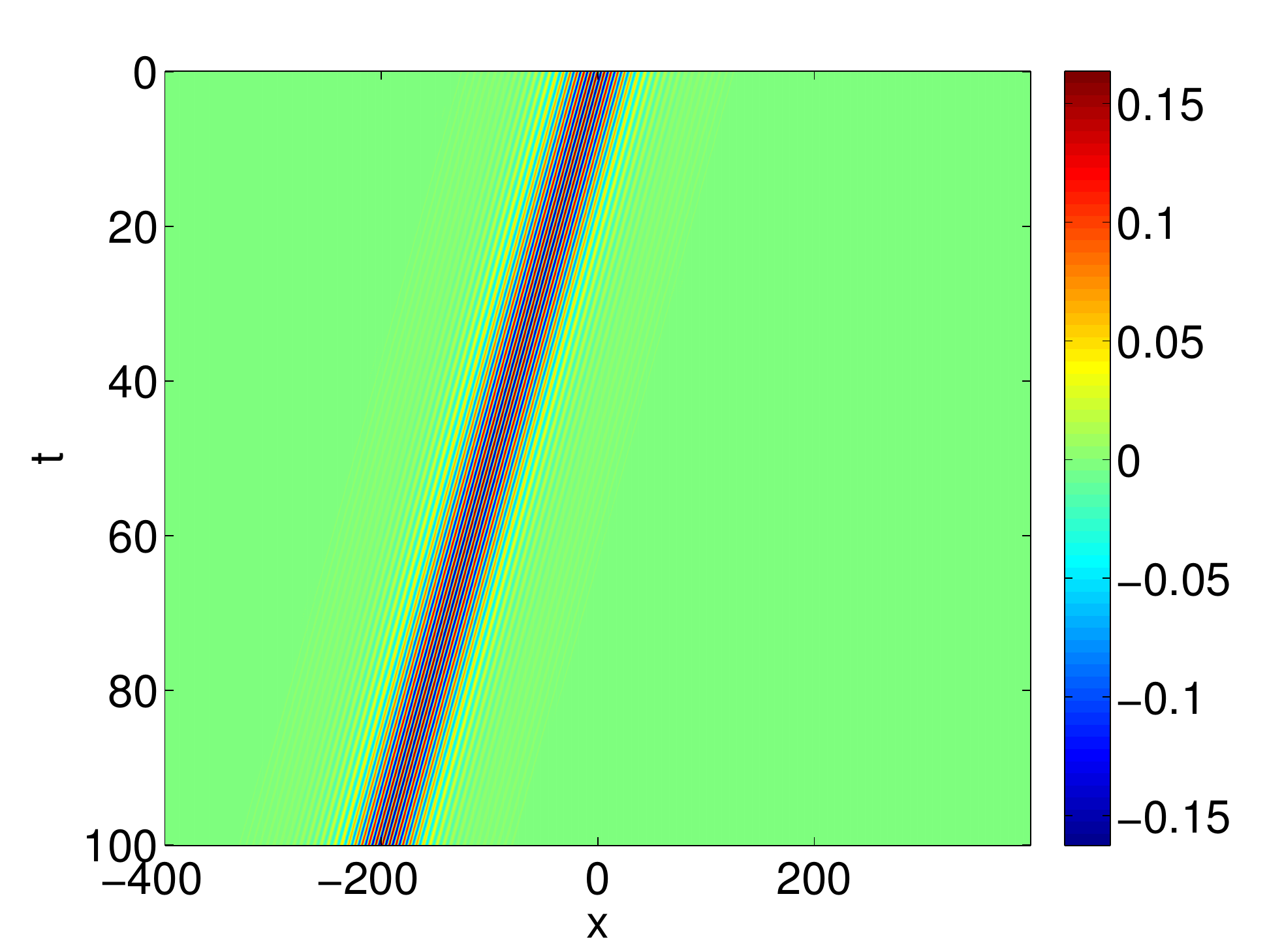}
	\includegraphics[width=.32\textwidth]{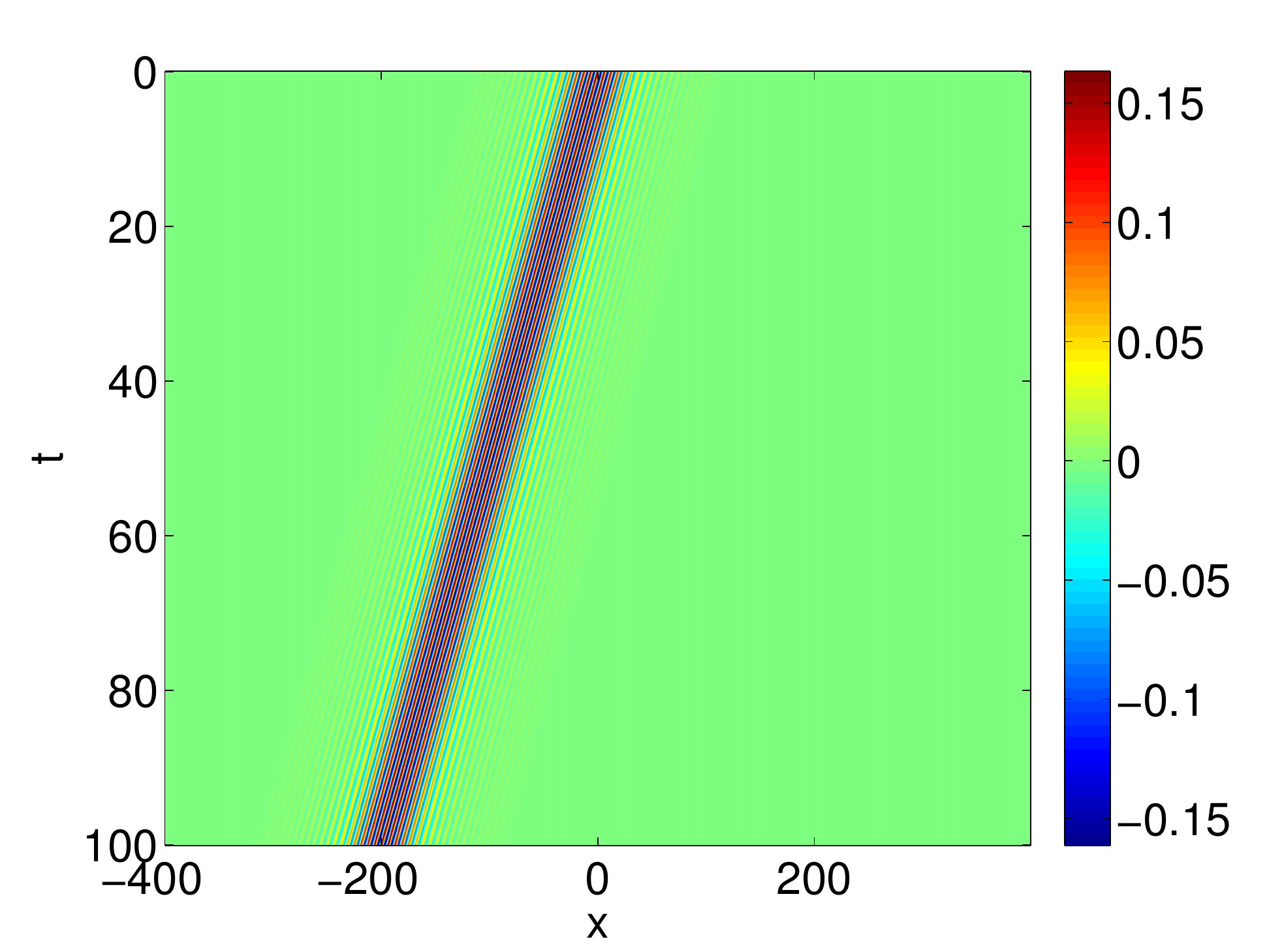}
	\includegraphics[width=.32\textwidth]{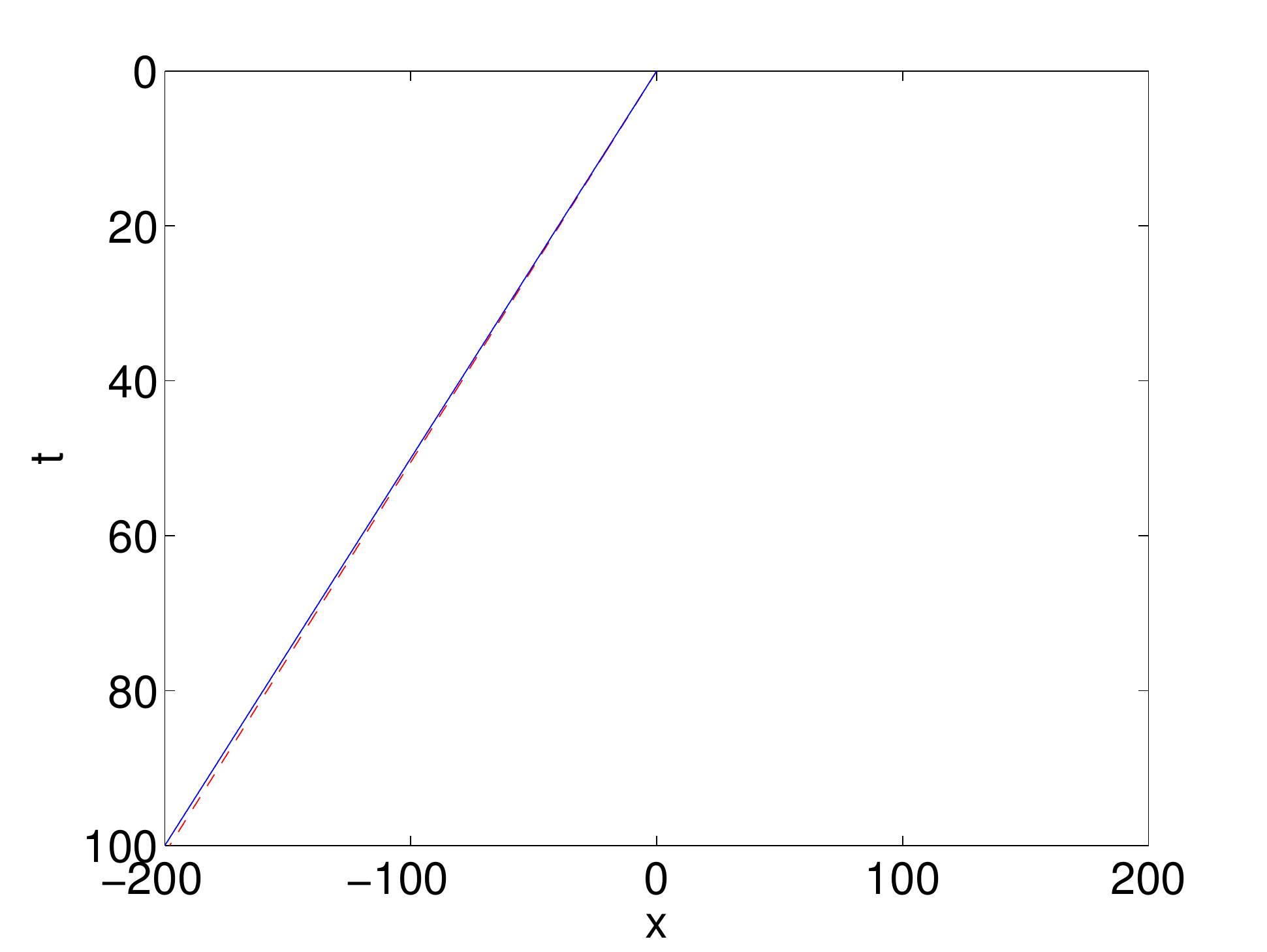}
	\caption{(Color online) Top panels:
	initial data as reconstructed from the NLS soliton (left), and a
	snapshot of the evolution at $t=100$;
	dashed (red) line represents the RSPE result
and the solid (blue) line the NLS prediction.
Bottom panels: evolution of the RSPE breather (left),
its theoretical NLS analogue (middle), and
center of mass evolution (right), using the same color notation as above.
Here, $\epsilon = 0.1$, $k=1$, $\alpha =1$, $\beta=1$, $c_g=-2$, $c_p = -2$.}
\label{a1_b1_k1}
\end{figure}
The above solution can directly be compared to respective solutions derived
in the framework of the RSPE. Note that, for the RSPE, the above NLS soliton
represents an approximate moving breather solution. Two typical such
examples, for a left moving and a right moving breather are shown in
Figs.~\ref{a1_b1_k1} and~\ref{a1_b0_k1}, respectively. The top panels show
the initial data (left-- common in both the ``theoretical'' NLS prediction
and in the RSPE) and the result of the evolution at $t=100$ (top right). The
agreement between the two, as illustrated by the field evolution contour
plots [bottom panels, RSPE (left) and NLS (middle)], and also by the center
of mass evolution (bottom right) clearly illustrates in a quantitative
fashion the accuracy of the multi-scale NLS-based approximation. For these
figures, the value $\epsilon=0.1$ has been used; notice, also, that in
Fig.~\ref{a1_b0_k1} refers to the case corresponding to the regular SPE limit
of $\beta=0$.

\begin{figure}[!htbp]
    \centering
	\includegraphics[width=.4\textwidth]{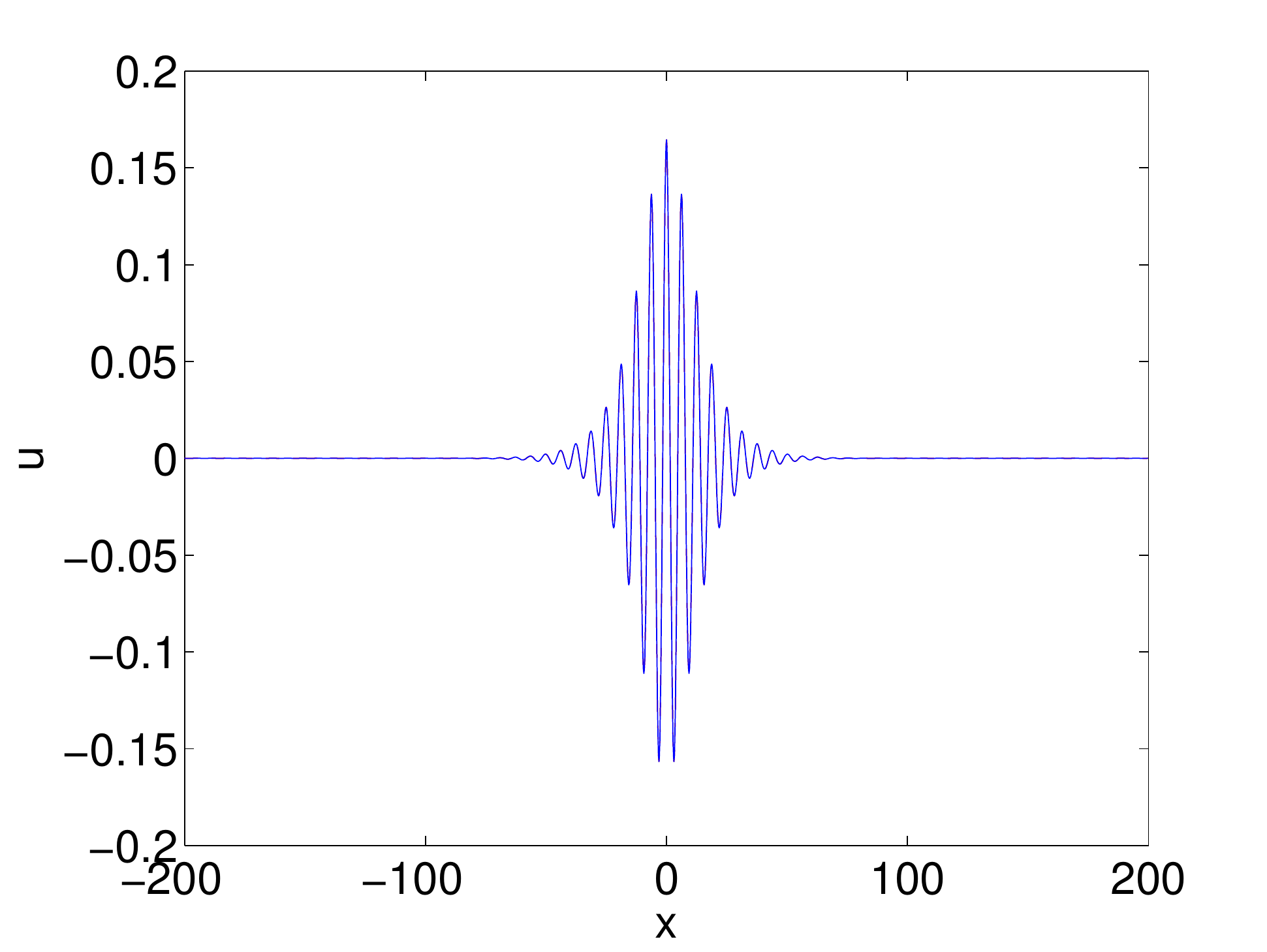}
	\includegraphics[width=.4\textwidth]{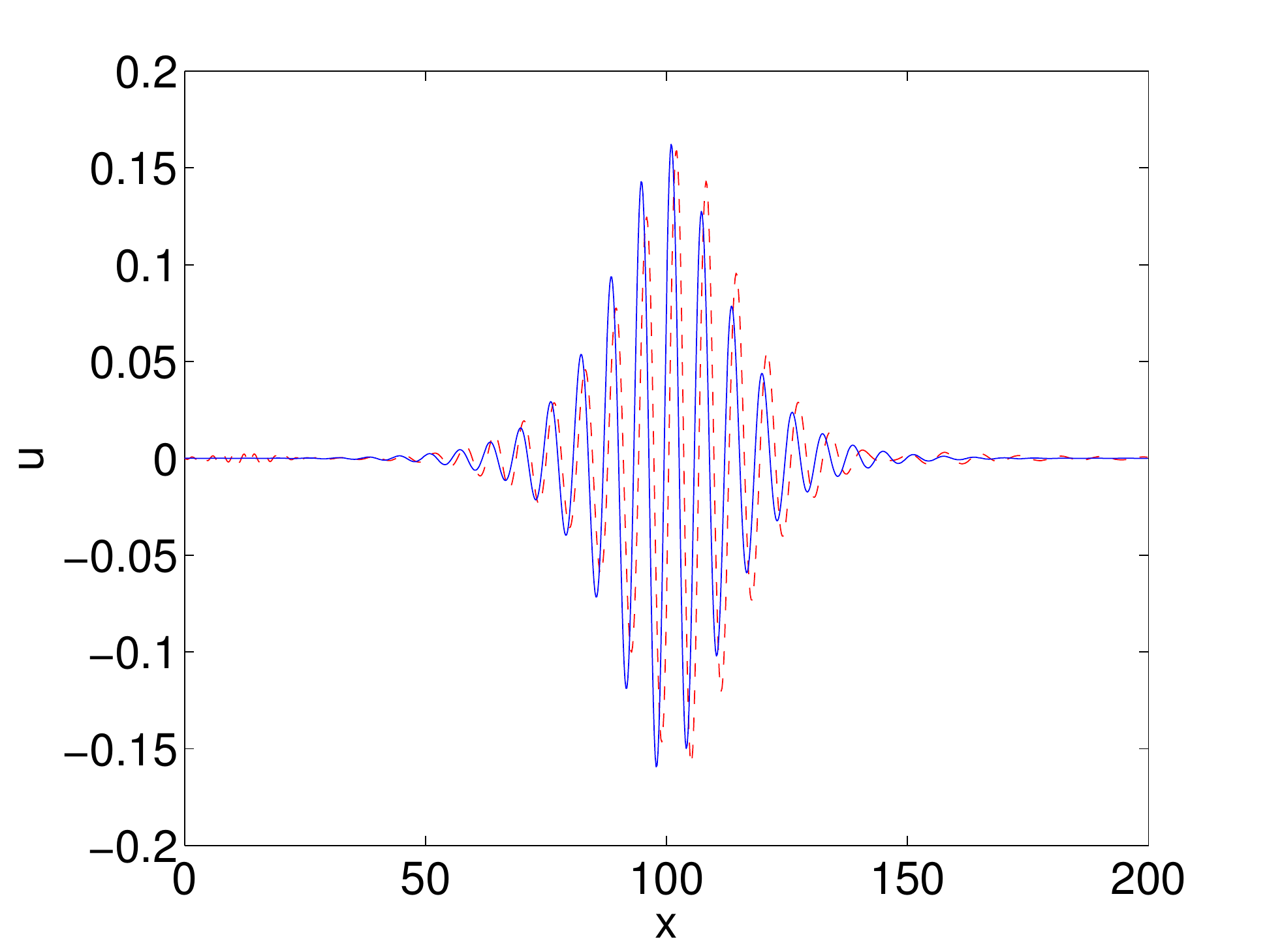}
	\includegraphics[width=.32\textwidth]{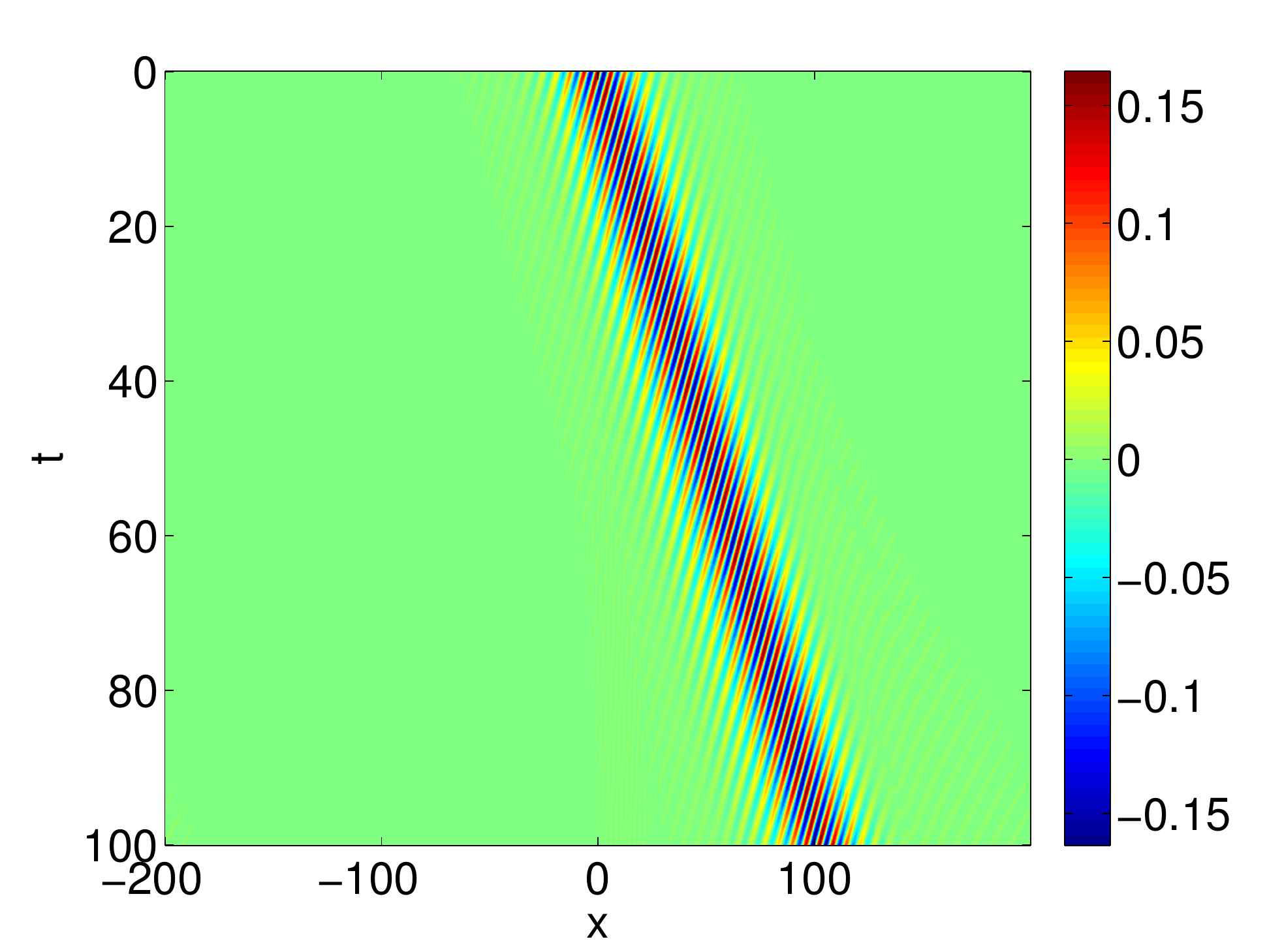}
	\includegraphics[width=.32\textwidth]{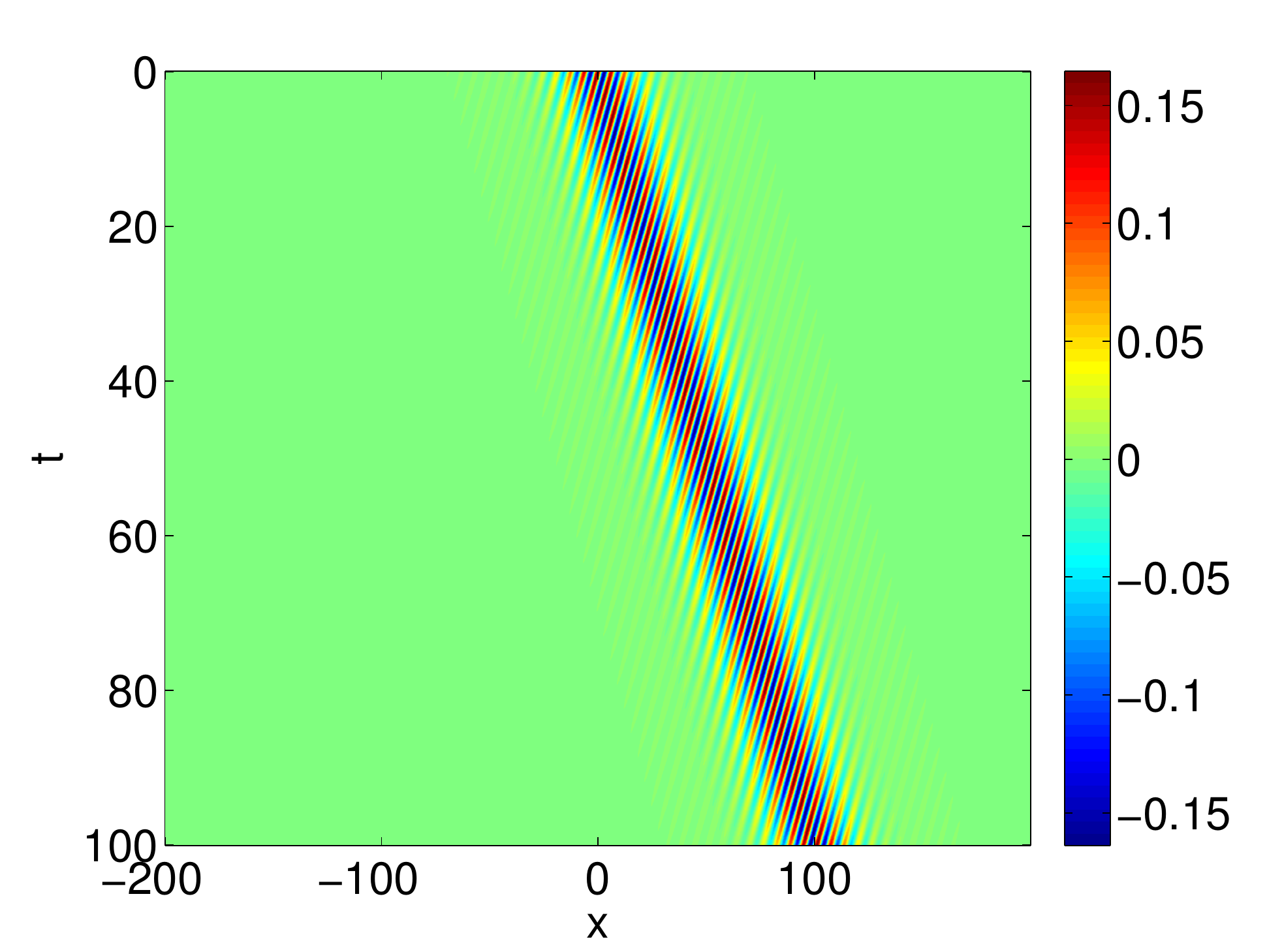}
	\includegraphics[width=.32\textwidth]{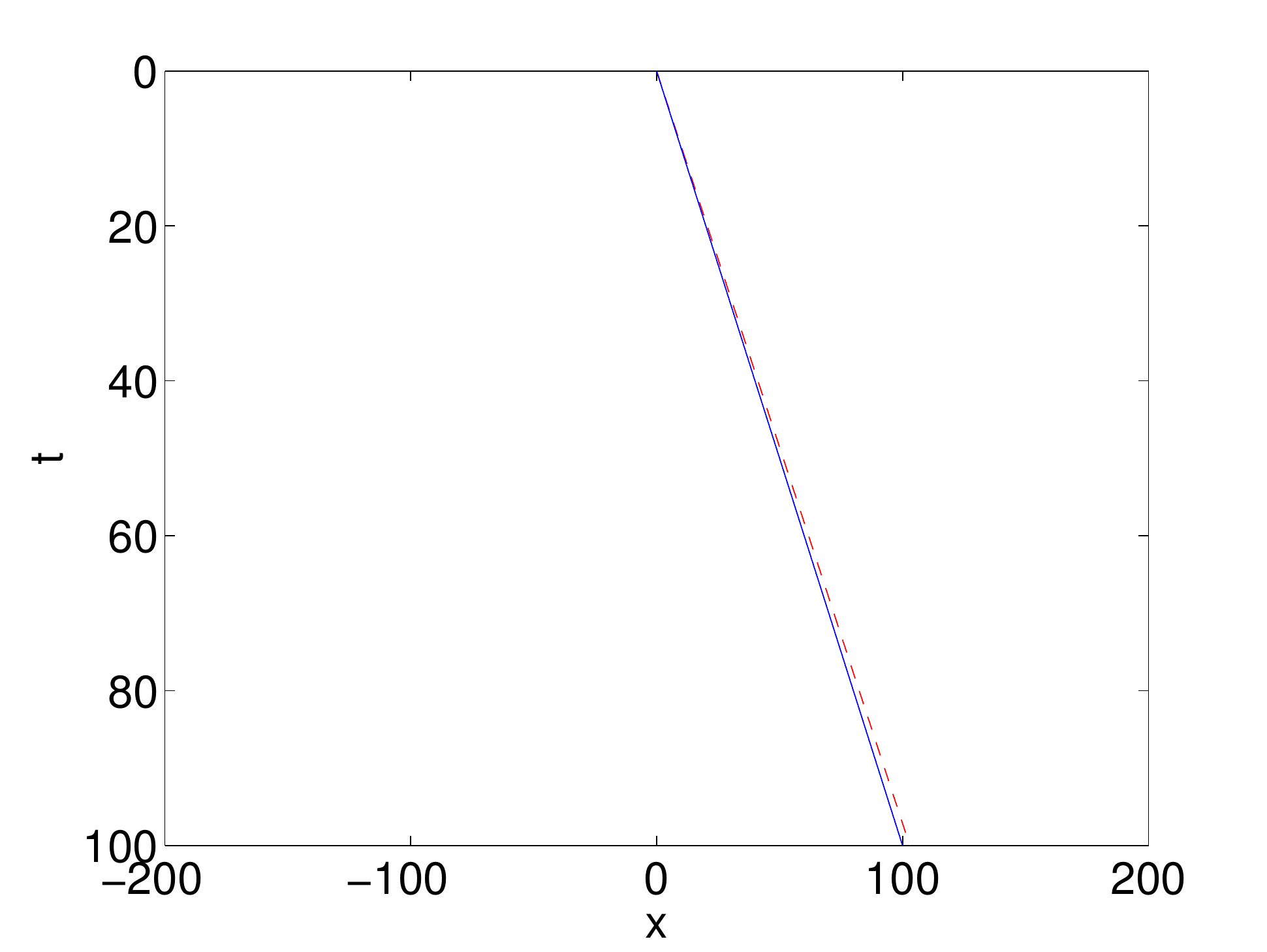}
		\caption{Same as Fig. \ref{a1_b1_k1}, but for $\epsilon = 0.1$, $k=1$, $\alpha =1$, $\beta=0$,
$c_g=1$, $c_p = -1$. Notice that the choice $\beta=0$ corresponds to the SPE (rather than the RSPE) model.}
\label{a1_b0_k1}
\end{figure}

\begin{figure}[!htbp]
    \centering
	\includegraphics[width=.4\textwidth]{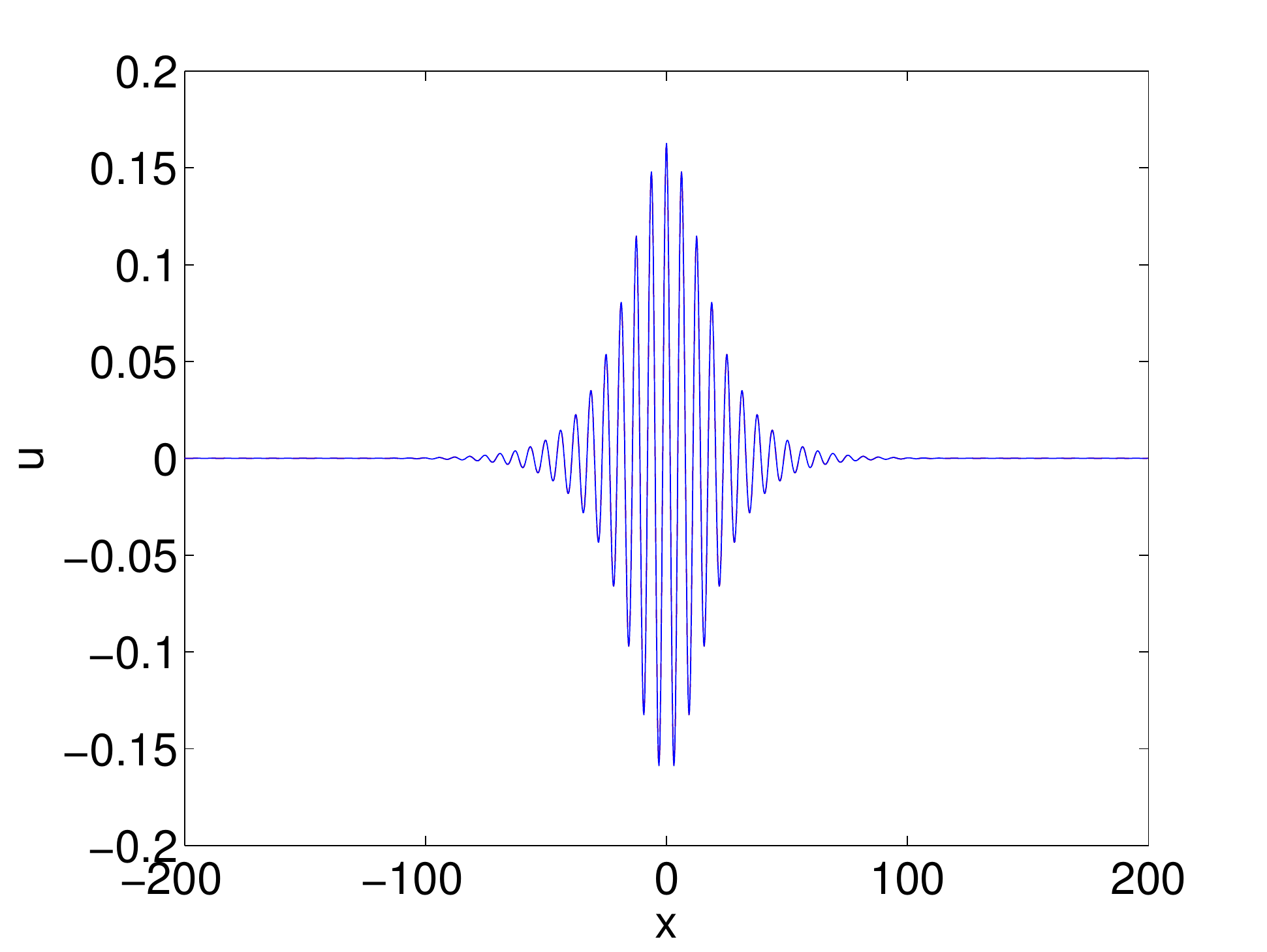}
	\includegraphics[width=.4\textwidth]{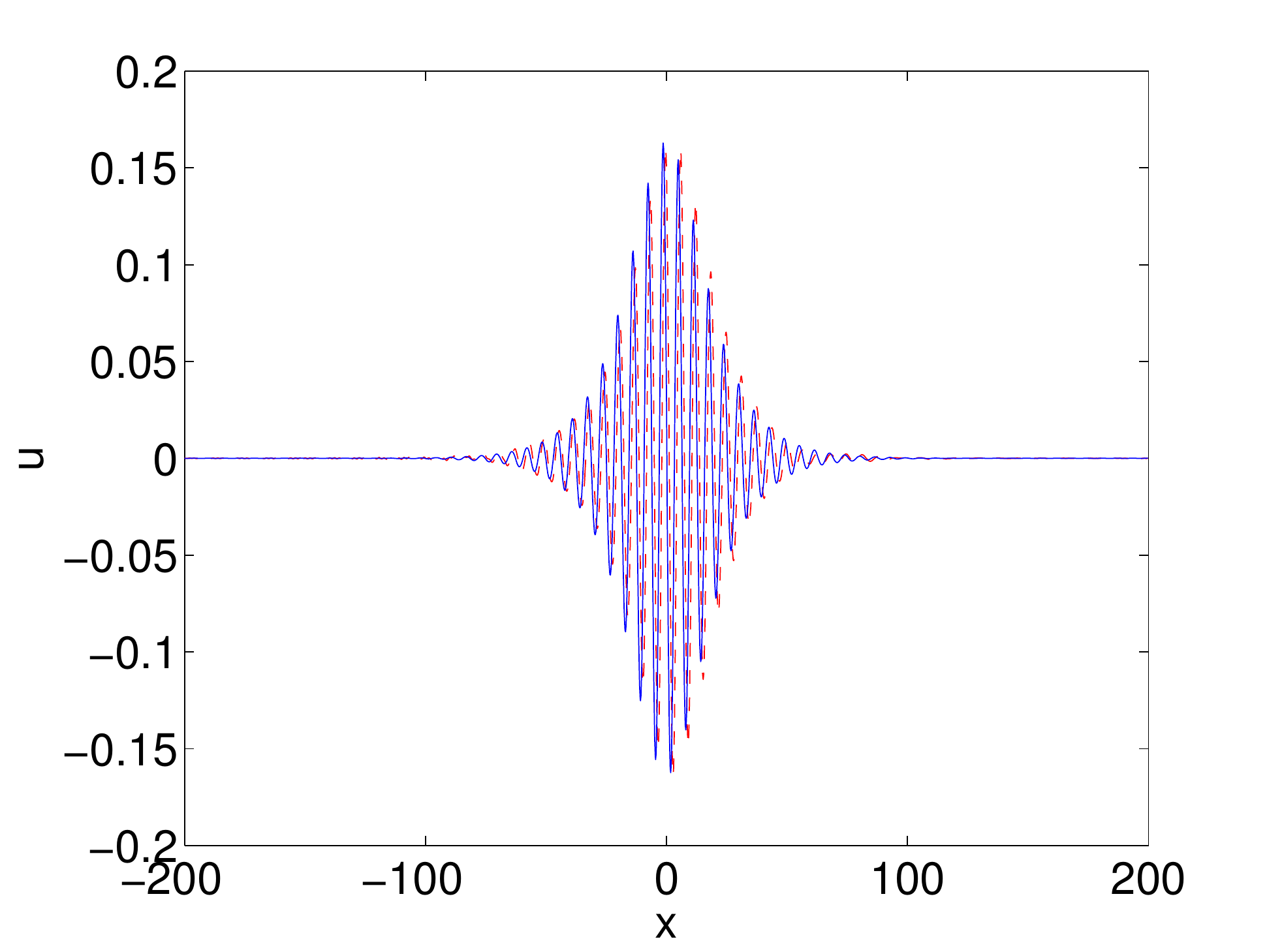}
	\includegraphics[width=.32\textwidth]{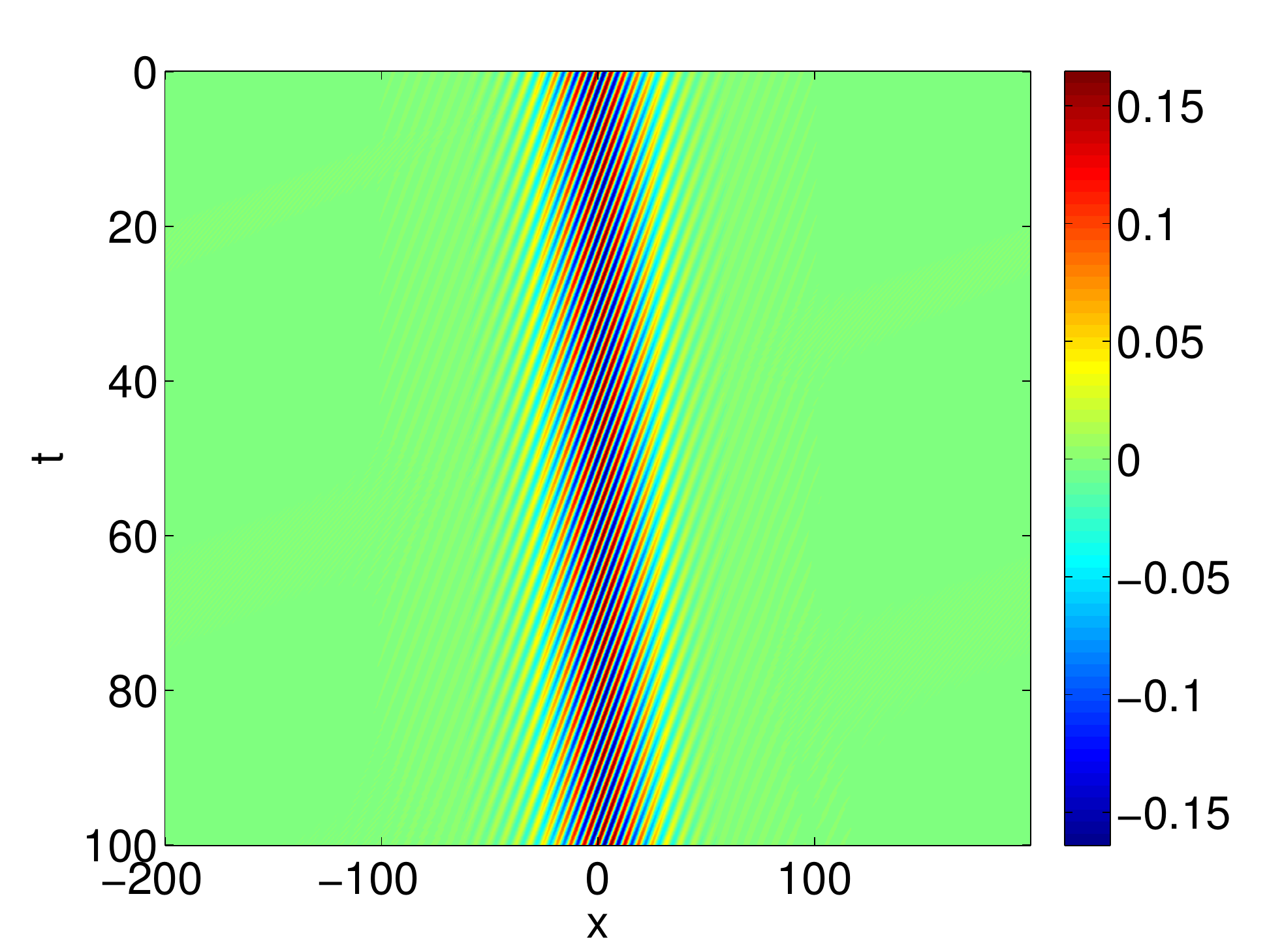}
	\includegraphics[width=.32\textwidth]{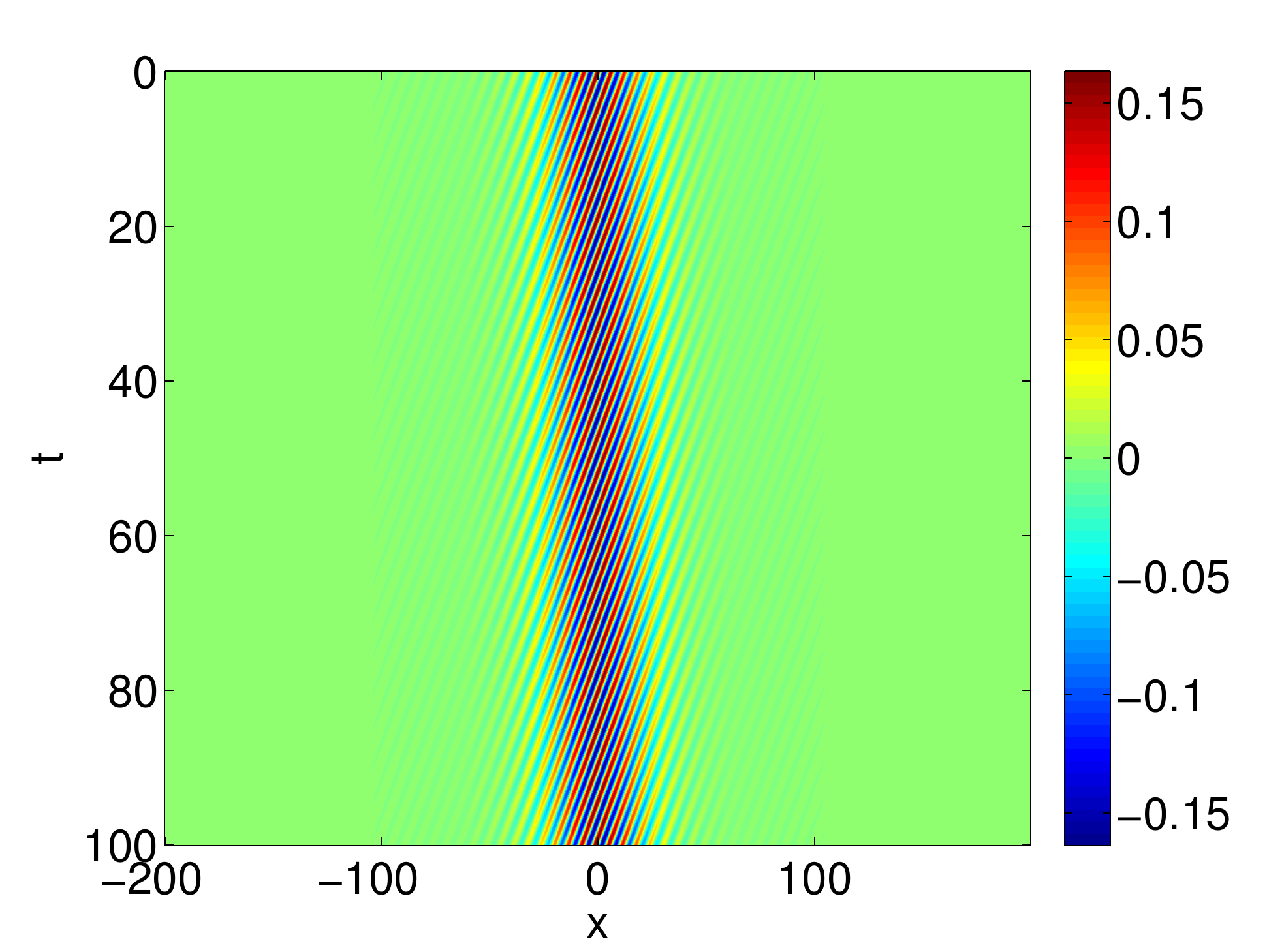}
	\includegraphics[width=.32\textwidth]{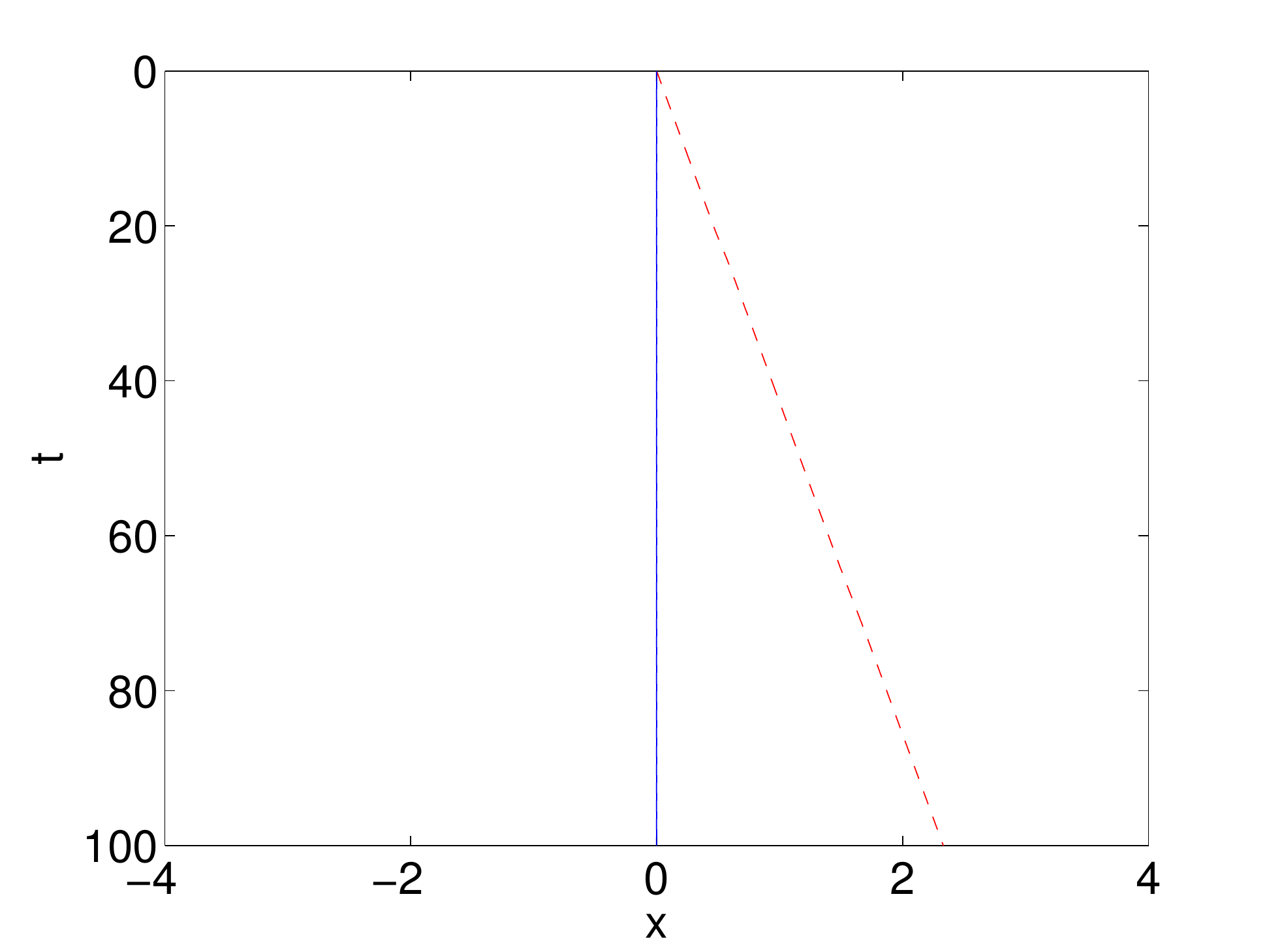}
		\caption{Same as Fig. \ref{a1_b1_k1}
but for $\epsilon = 0.1$, $k=1$, $\alpha =1$, $\beta=\frac{1}{3}$, $c_g=0$, $c_p = -\frac{4}{3}$. }
\label{a1_b03d_k1}
\end{figure}

\begin{figure}[!htbp]
    \centering
	\includegraphics[width=.4\textwidth]{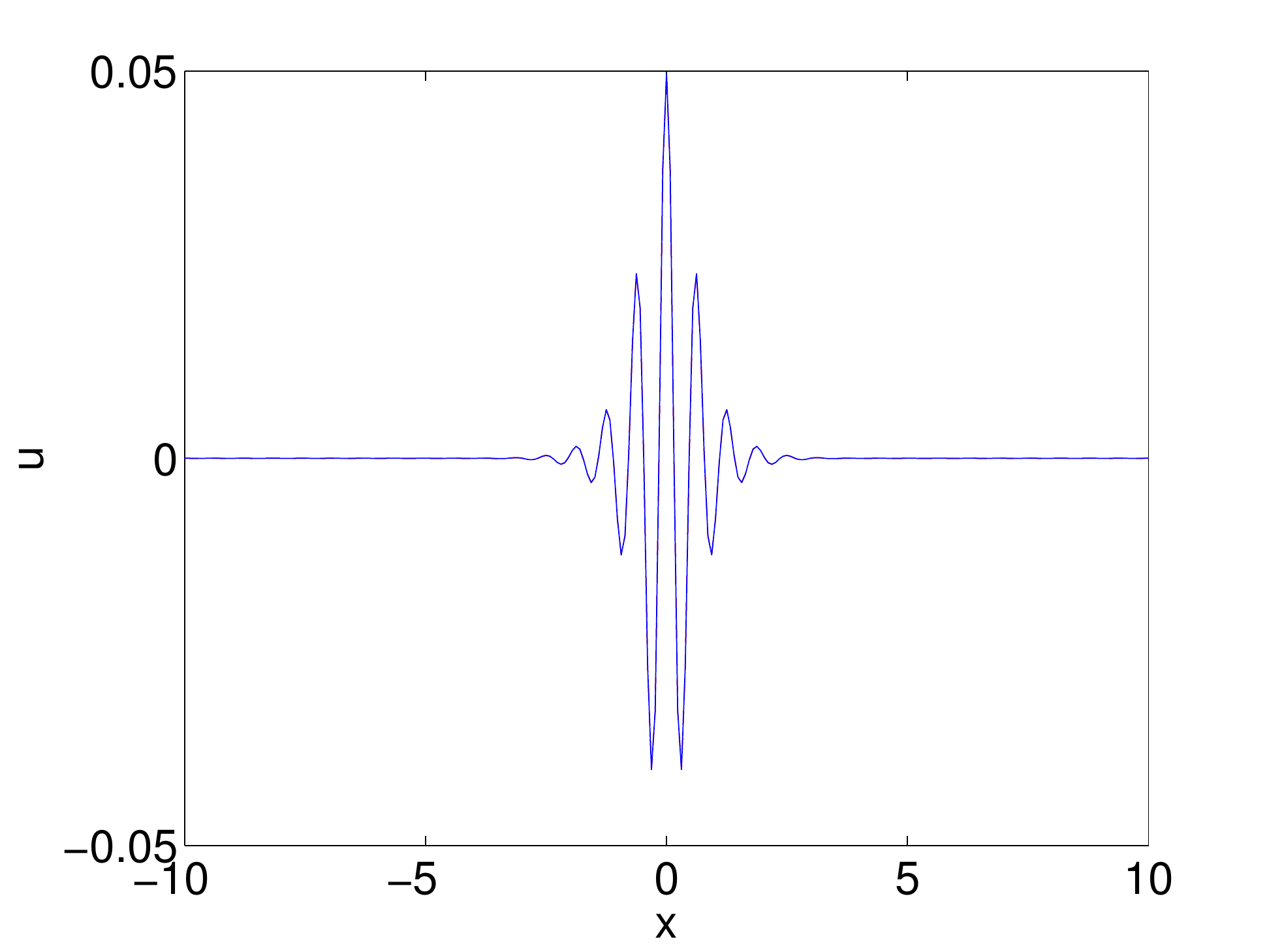}
	\includegraphics[width=.4\textwidth]{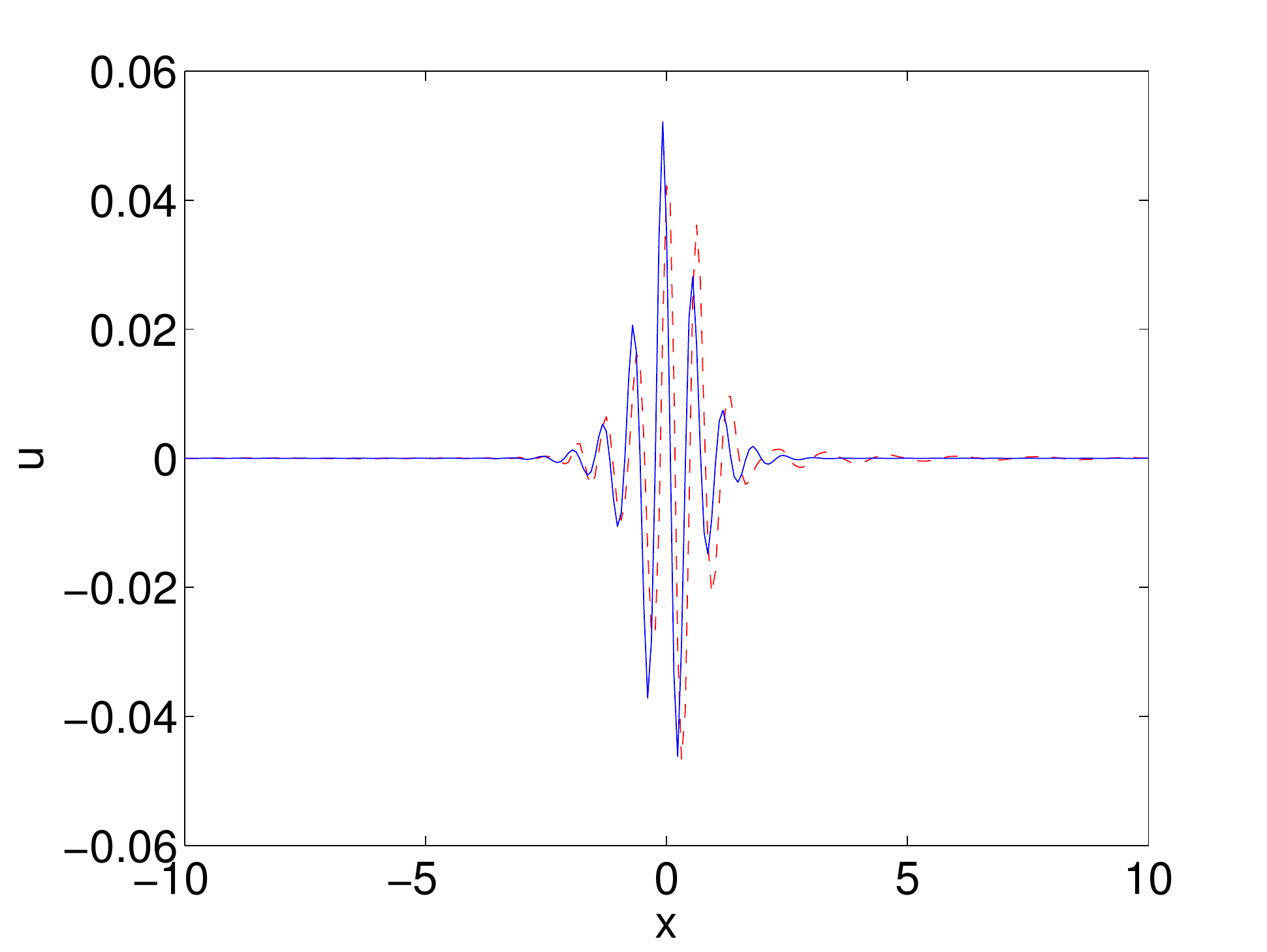}
	\includegraphics[width=.32\textwidth]{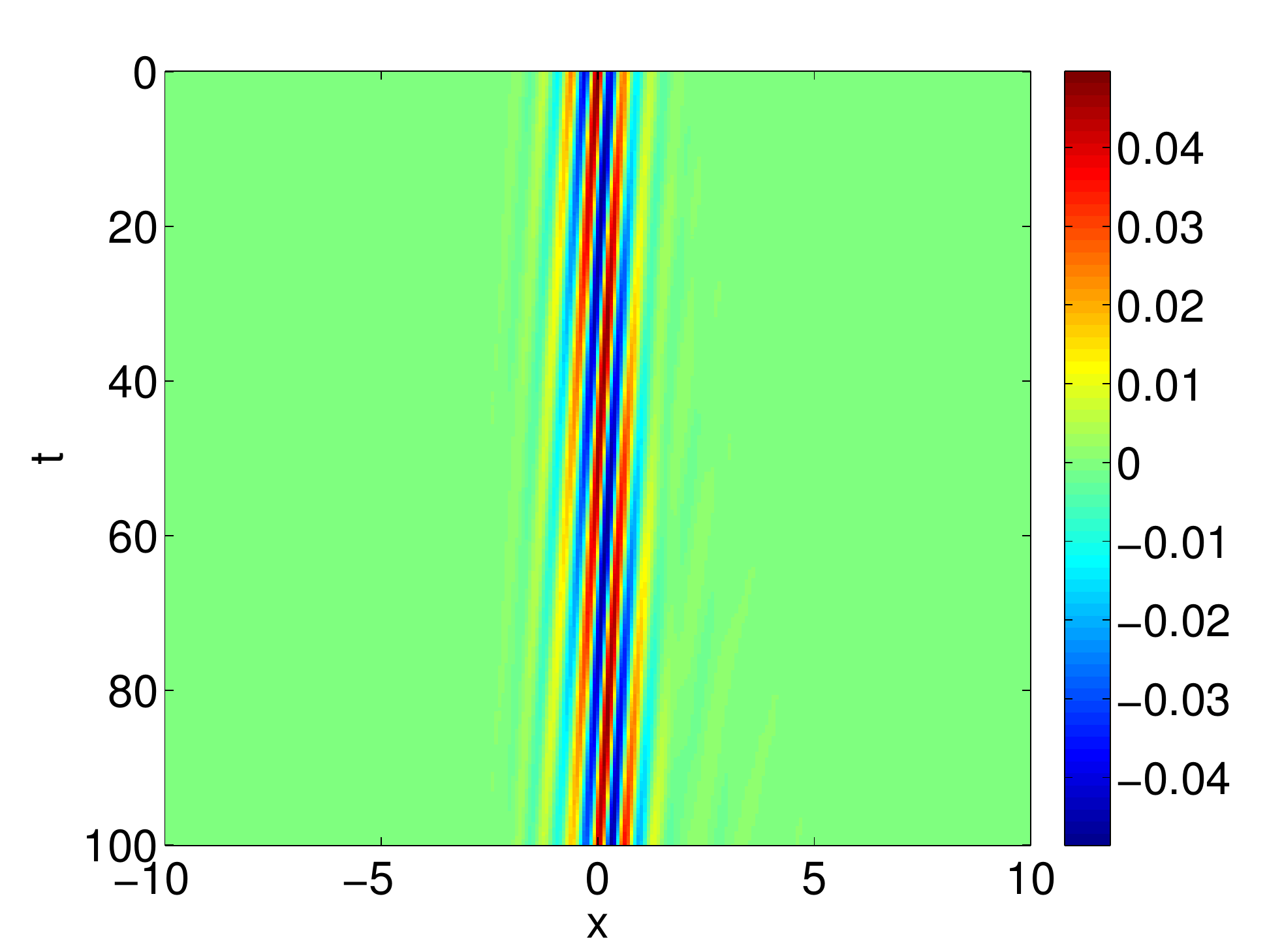}
	\includegraphics[width=.32\textwidth]{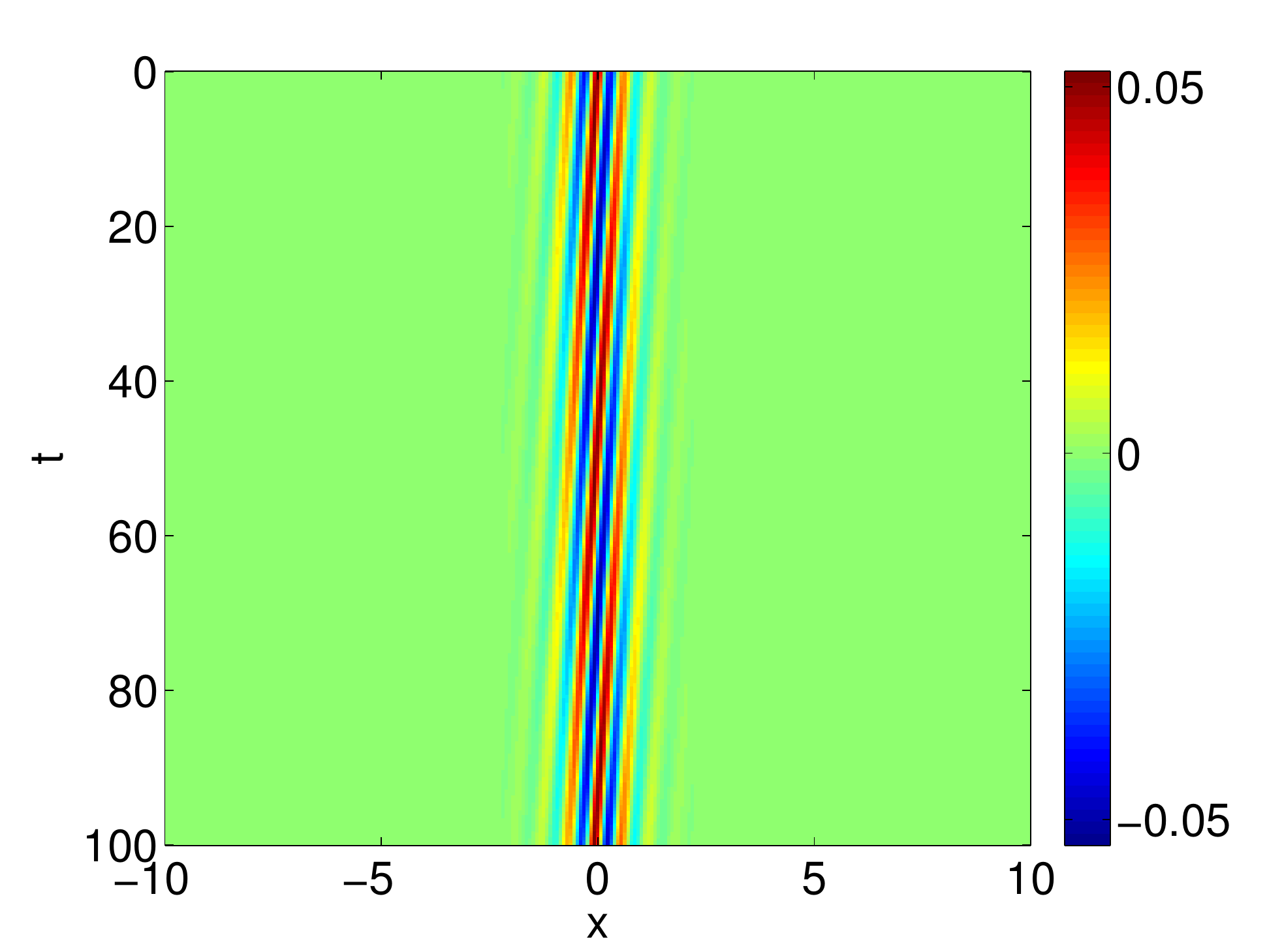}
	\includegraphics[width=.32\textwidth]{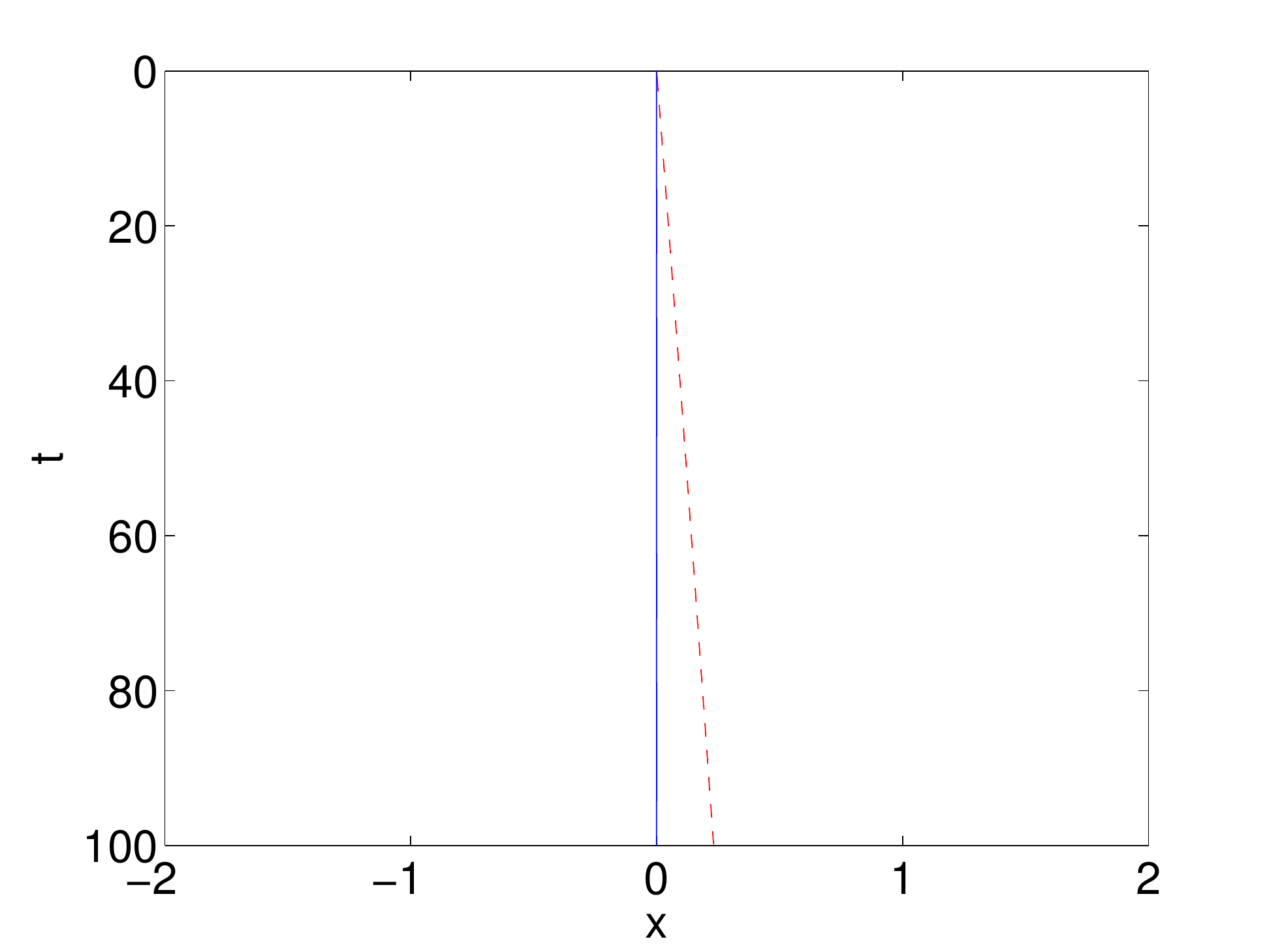}
	\caption{Same as Fig. \ref{a1_b1_k1} but for $\epsilon = 0.1$,
	$k=10$, $\alpha =1$, $\beta=\frac{1}{30000}$, $c_g=0$, $c_p =-\frac{4}{300}$. }
	\label{a1_b000003d_k10}
\end{figure}

A more demanding comparison is shown in Fig.~\ref{a1_b03d_k1} and also in
Fig.~\ref{a1_b000003d_k10} (for a finite $\beta$ of $1/3$ and a near-zero
value of $\beta$, i.e., the SPE limit, respectively). In these cases, the
breather is theoretically expected to be stationary. However, as observed in
both figures (i.e., both in the RSPE and in the SPE case), the breather
slowly drifts, over long time scales, over short distances away from its
original position. Nevertheless, this drift is very slow and is nearly
imperceptible over the propagation distances shown in the figure, especially
so in Fig.~\ref{a1_b000003d_k10}.

From all of the above cases, it can be readily inferred that the breather
structures that were predicted to be robust in the SPE, are also found to be
equally robust in the context of the RSPE in all the cases tested (including
ones not shown here). Such breathers can be constructed fairly accurately, at
least for sufficiently small amplitudes, using NLS bright solitons as
suitable time and space modulated constituents.

\subsection{Dark Breathers}

The formal derivation of the NLS equation from the RSPE model allows for the
prediction of still another approximate solution of the RSPE. Indeed, as mentioned in
Section~3.1, the NLS model also admits dark soliton solutions, for $PQ<0$. Such a solution
is of the form:
\[
A(\tau,\xi) = \sqrt{|Q|}\tanh\left(\frac{\xi}{\sqrt{2|P|}}\right)\exp(-i\tau).
\]
Formally, and in the framework of the RSPE, such solutions exist only for
$\beta<0$ [recall that bright breathers do not exist in this case --cf.
Eq.~(\ref{cond})]. In addition, since such solutions of the NLS exhibit
nonvanishing boundary conditions at $\xi \rightarrow \pm \infty$, it is clear
that they cannot be supported as approximate solutions of the RSPE on the
line ($-\infty <x <+\infty$), as the only uniform stationary state of the
latter is $u=0$. Nevertheless, since the field $u(x,t)$ obeying the RSPE
model is real, it is possible to construct approximate dark breather
solutions of the RSPE (based on the dark solitons of the NLS), using periodic
boundary conditions.

\begin{figure}[!htbp]
    \centering
	\includegraphics[width=.32\textwidth]{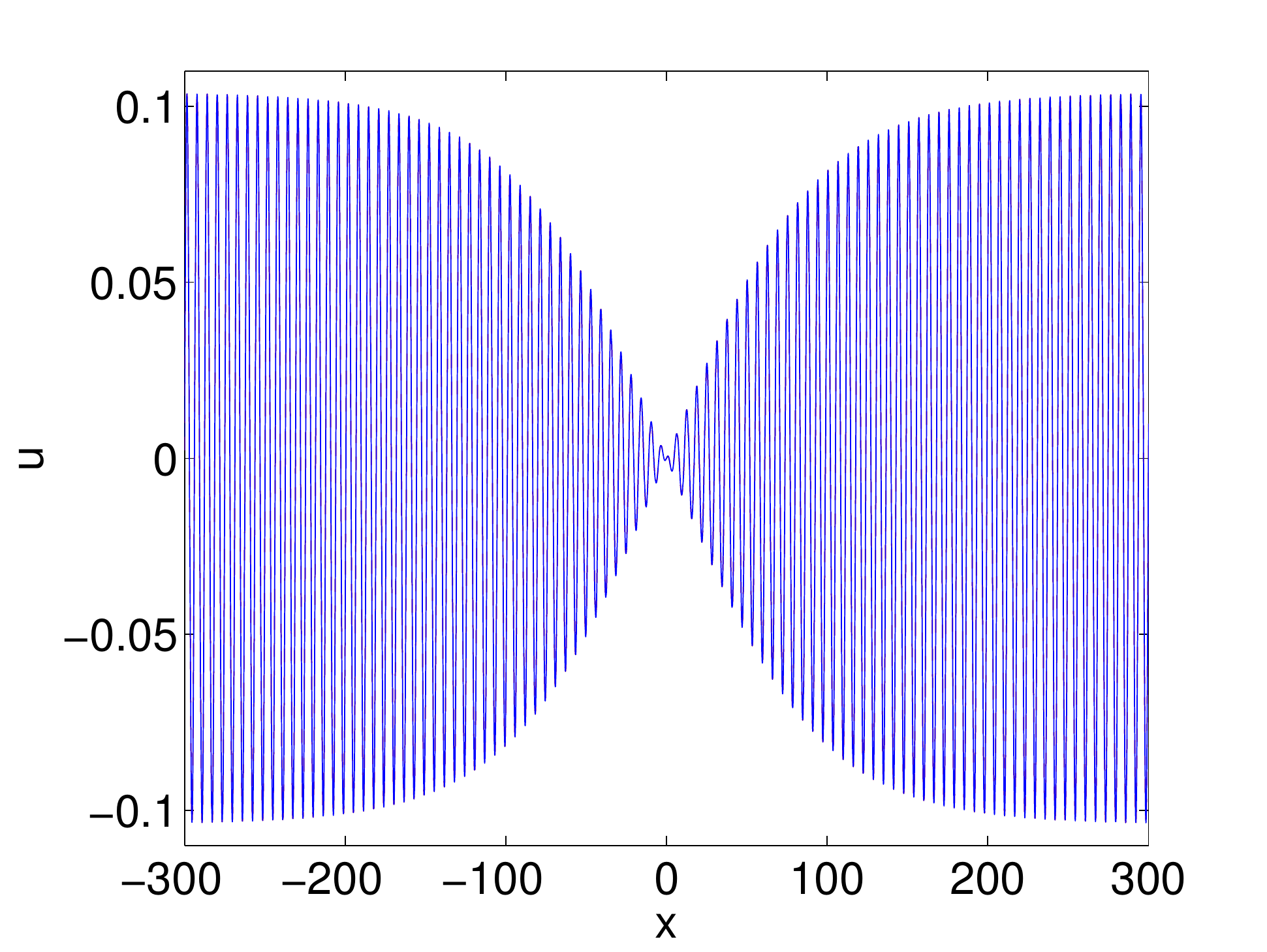}
	\includegraphics[width=.32\textwidth]{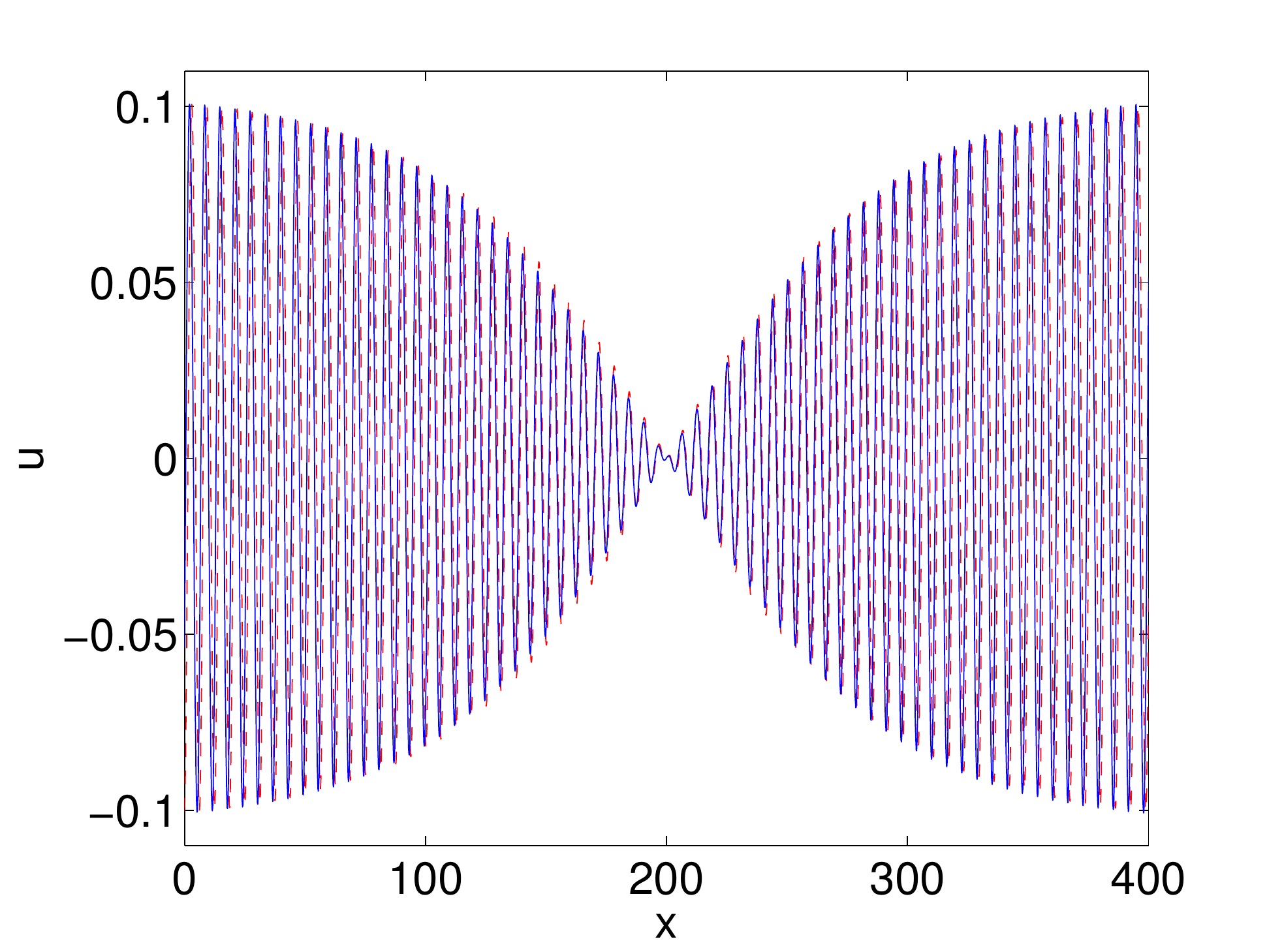}
	\includegraphics[width=.32\textwidth]{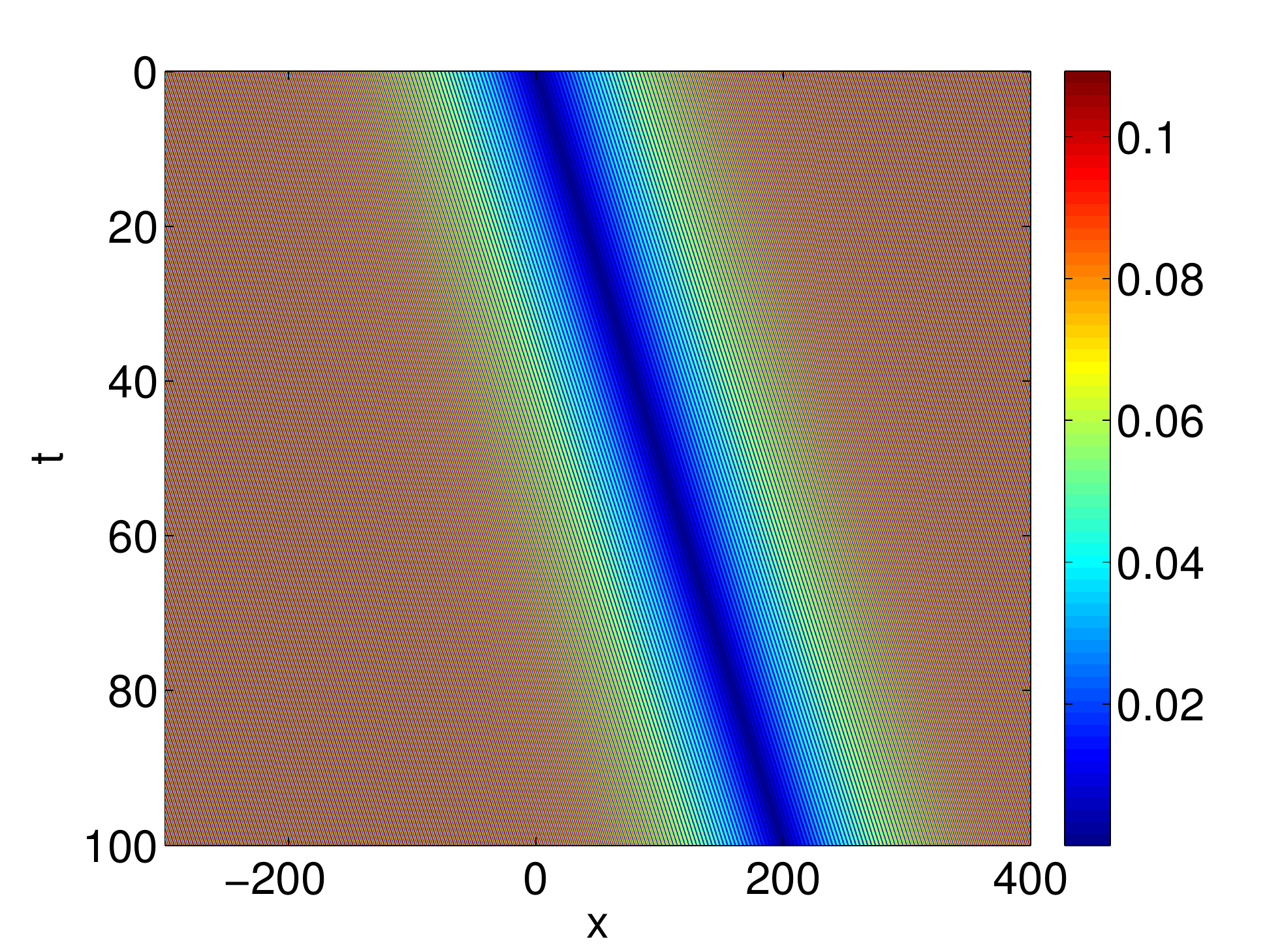}
		\caption{(Color online) 
		The dark soliton at $t=0$ (left panel) and at
$t=100$ (middle panel); again the dashed (red) line is for RSPE and solid (blue) line is for the NLS
approximation.
The contour plot (right panel) shows
the space-time evolution of this approximate solution under the RSPE model. Here, $\epsilon = 0.03$, $k=1$,
$\alpha =-1$, $\beta=-1$, $c_g=2$, $c_p =2$.}
\label{dark}
\end{figure}

Such an approximate dark breather of the RSPE is shown in Fig.~\ref{dark}. It
is clearly observed that, once again, the NLS approximation is excellent
(left and middle panels), since the coherent structure propagates undistorted
over many cycles during the course of the simulation.

\section{Conclusions \& Future Challenges}

In the present work, we have revisited the regularized short pulse equation
(RSPE) introduced in the earlier works of~\cite{cmj,mcjs}. We have examined
two prototypical classes of solutions. The first one is the traveling wave
(rigorously predicted to occur in a single- and multi-pulse form
in~\cite{cmj,mcjs}). The second solution stems from the standard short pulse
equation (SPE) without the regularization, and has the form of a breather.

For the traveling solitary waves of the equation, we used a fixed point
iteration scheme to obtain these solutions for different values of the RSPE
coefficients. These solutions are found in our numerical evolution
simulations to be robust and maintain their characteristics even in the
presence of noise. It was found that they can be well approximated by a
truncated series of hyperbolic secants.

As concerns the breather solutions, they were found by means of a multi-scale
expansion reducing the RSPE to the nonlinear Schr{\"o}dinger (NLS) equation.
From the standard sech-soliton solution of the latter, the 
bright breather of the
RSPE was systematically reconstructed and its robustness was again shown by
means of direct numerical simulations. Furthermore, using the dark
(tanh-shaped) soliton solution of the NLS, we constructed approximate dark
breather solutions of the RSPE model using periodic boundary conditions.
These solutions, which exist for parameter values where regular (bright)
breathers are not supported by the system, were found to be robust in the
simulations as well.

There are many directions that one can envision in terms of possibilities for
future work based on the findings of the present study. On the one hand, it
would be relevant to attempt to analyze the form and stability of multi-pulse
solutions both through numerical computations and through rigorous analysis
(extending further the work of~\cite{mcjs}). On the other hand, generalizing
the RSPE model to higher dimensions, in the spirit of Ref.~\cite{shen2d},
would be of interest in its own right. Then, it would be relevant to explore
the 2D generalization of the reductions proposed and solutions analyzed
herein to examine whether such breathers, and especially genuinely 2D
traveling waves, could exist in such a model. Furthermore, of particular
interest may be the study of rogue waves under the RSPE and its water waves
counterpart, namely the Ostrovsky equation~\cite{ostro}. Such studies are
currently under way and will be reported in future publications.

\section*{Acknowledgements.} The work of D.J.F. was partially supported
by the Special
Account for Research Grants of the University of Athens.
PGK acknowledges support from the
Alexander von Humboldt Foundation, the Binational
Science Foundation under grant 2010239, NSF-CMMI-1000337,
NSF-DMS-1312856, FP7, Marie Curie Actions, People, International
Research Staff Exchange Scheme (IRSES-606096)
and from the US-AFOSR under grant FA9550-12-10332.

\appendix

\section{Spectral renormalization for the RSPE}

The idea behind the  spectral renormalization  method is to transform the
underlying equation governing the soliton into Fourier space and determine a
nonlinear nonlocal integral equation coupled to an algebraic equation. This
coupling generally prevents  the numerical scheme from diverging. In
particular, seeking traveling wave solutions of the RSPE, Eq.~(\ref{rspe}),
we start our analysis by considering Eq.~(\ref{rspe2}) with the boundary
conditions
\[
u\rightarrow 0\quad \mathrm{as}\quad |\xi|\rightarrow \infty.
\]
Now define the Fourier transform (FT) pair as
\begin{eqnarray*}
\hat{u}(\omega)&=\mathcal{F}\{ u(\xi) \}=\int_{-\infty}^{\infty}
u(\xi)\;e^{-i\omega \xi}\;d\xi, \\
u(\xi) &= \mathcal{F}^{-1}\{ \hat{u}(\omega) \}=\frac{1}{2\pi} \int_{-\infty}^{\infty}
\hat{u}(\omega)\;e^{i\omega \xi}\;d\omega.
\end{eqnarray*}
Applying the FT to Eq.~(\ref{rspe2}) we obtain the equation:
\[
(\alpha+c\omega^2+\beta\omega^4)\hat{u}-\omega^2\mathcal{F}\{ u^3 \}=0.
\]
In order to construct a solution so that its amplitude does not grow
indefinitely nor tends to zero with each iteration, we introduce $v(\xi)$
such that:
\[
u(\xi)=\lambda v(\xi) \Leftrightarrow
\hat{u}(\omega)=\lambda \hat{v}(\omega),
\]
where $\lambda$ is to be determined. Then $v(\xi)$ satisfies:
\[
(\alpha+c\omega^2+\beta\omega^4)\hat{v}-\lambda^2\omega^2\mathcal{F}\{ v^3 \}=0.
\]
Multiplying by $\hat{v}^*(\omega)$ and integrating over the entire space
$\omega$ we find the relation:
\[
\int_{-\infty}^{\infty} (\alpha+c\omega^2+\beta\omega^4)|\hat{v}|^2\; d\omega- \lambda^2
\int_{-\infty}^{\infty} \omega^2\mathcal{F}\{ v^3 \}\hat{v}^*\;d\omega=0,
\]
which must then be solved for $\lambda$, thus determining the necessary
constant so that the solution will not simply decay to zero or blow-up after
an iteration. Finally, the solution is obtained by iterating as follows:
\[
\hat{v}_{n+1}(\omega)=\frac{\lambda_n^2\omega^2\mathcal{F}\{ v_n^3 \}}{(\alpha+c\omega^2+\beta\omega^4)}.
\]
Recall, that $\alpha$, $\beta$ and $c$ share the same sign so the denominator
has no zeros. This iteration is for $n>0$. When $n=0$ an initial guess is
required, typically a Gaussian.

To ensure that convergence is achieved we employ in our numerical codes three
criteria. We demand convergence in $\lambda$, namely
\[
|\lambda_{n+1}-\lambda_n|<\delta,
\]
convergence in $v$,
\[
|v_{n+1}-v_n|<\delta,
\]
where $\delta$ is the accuracy we desire, say $10^{-9}$ or $10^{-10}$, and
most importantly when these two conditions are satisfied the solution
satisfies the equation up to the same accuracy $\delta$, or in other words,
the residual is of that order.

\section{The higher-order NLS equation and its soliton solutions}

The multiscale analysis presented in Sec.~3.1 leads to the regular NLS model,
Eq.~(\ref{NLS}), at the order $O(\epsilon^3)$. At the next order of approximation,
namely at $O(\epsilon^4)$, it is possible to derive a higher-order NLS, which
incorporates third-order dispersion and higher-order nonlinear terms. This
model is of the form,
\begin{eqnarray}
i\partial_{\tau}A +P\partial_{\xi}^2 A + Q|A|^2A+ i\delta \partial_{\xi}^3A
+i\mu |A|^2 \partial_{\xi} A + i\nu A^2 \partial_{\xi} A^*=0,
\label{hoNLS}
\end{eqnarray}
where the coefficients of the higher-order terms are given by:
\[
\delta=4\beta \epsilon, \qquad \mu = 12\epsilon, \qquad \nu=6\epsilon.
\]
Following the analysis of Ref.~\cite{old}, it is possible to derive exact
travelling wave solutions of Eq.~(\ref{hoNLS}), in the form of solitary waves.
Such solutions are sought in the form:
\[
A(\xi,\tau)= \Phi(\eta)\exp[i(K\xi-\Omega \tau - \theta)], \qquad \eta = \xi-V\tau,
\]
where $\Phi(\eta)$ is an unknown real function, while $K$, $\Omega$ and $V$ denote
the wavenumber, angular frequency and velocity of the wave ($\theta$ is a constant phase).
Introducing the above ansatz
into Eq.~(\ref{hoNLS}), and separating real and imaginary parts, we derive the following
ordinary differential equations (ODEs):
\begin{eqnarray}
&&(P-3\delta K)\Phi'' +(\Omega-P K^2+\delta K^3)\Phi +[Q-(\mu+\nu)K]\Phi^3=0,
\label{s1} \\
&&\delta \Phi''' +(-V+2PK-3\delta K^2)\Phi' +(\mu+\nu)\Phi^2 \Phi'=0,
\label{s2}
\end{eqnarray}
where primes denote derivatives with respect to $\eta$. Observing that the compatibility
condition of the above equations (which is found upon differentiating Eq.~(\ref{s1}) once
with respect to $\eta$) is
\begin{eqnarray}
&&\frac{\Omega-P K^2+\delta K^3}{P-3\delta K} = \frac{-V+2PK-3\delta K^2}{\delta}=\lambda_1,
\label{p1}
\\
&&\frac{3[Q-(\mu+\nu)K]}{P-3 \delta K} = \frac{\mu+\nu}{\delta}=\lambda_2,
\label{p2}
\end{eqnarray}
we find that Eqs.~(\ref{s1}), (\ref{s2}) are equivalent to the following ODE:
\begin{equation}
\Phi'' +\lambda_1 \Phi + \lambda_2 \Phi^3=0.
\label{duf}
\end{equation}
Then, utilizing Eq.~(\ref{duf}), it is straightforward to show (see, e.g., Ref.~\cite{old} for details),
that there exist bright and dark solitary wave solutions of Eq.~(\ref{hoNLS}). In particular,
in the case $\lambda_1<0$ and $\lambda_2>0$, we derive the bright solitary wave
\begin{equation}
\Phi(\eta)= \left(\frac{2|\lambda_1|}{\lambda_2}\right)^{1/2}{\rm sech}(\sqrt{|\lambda_1|}\eta),
\label{bs}
\end{equation}
while in the case $\lambda_1>0$ and $\lambda_2<0$, we derive the dark solitary wave
\begin{equation}
\Phi(\eta)=
\left(\frac{\lambda_1}{|\lambda_2|}\right)^{1/2}\tanh\left(\sqrt{\frac{|\lambda_2|}{2}}\eta\right).
\label{ds}
\end{equation}

Observing that the three parameters $K$, $\Omega$ and $V$ of the above solutions
are connected via two equations, (\ref{p1})-(\ref{p2}), it is clear that the solitary waves
in Eqs.~(\ref{bs})-(\ref{ds})
are characterized by one free parameter.  Fixing this parameter,
we can appropriately fix the sign of $\lambda_1$; then, taking into regard that
$\lambda_2=(\mu+\nu)/2=(9/2)\beta$, we find that bright (dark) solitary waves for $\beta>0$ ($\beta<0$),
consistently
with the results obtained in the framework of the regular NLS model.

\end{document}